# Constructor Theory of Probability

Chiara Marletto[1]

Department of Materials, University of Oxford

April 2016

Unitary quantum theory, having no Born Rule, is *non-probabilistic*. Hence the notorious problem of reconciling it with the *unpredictability* and *appearance of stochasticity* in quantum measurements. Generalising and improving upon the so-called 'decision-theoretic approach' (Deutsch, 1999; Wallace, 2003, 2007, 2012), I shall recast that problem in the recently proposed *constructor theory of information* – where quantum theory is represented as one of a class of *superinformation theories*, which are *local*, *non-probabilistic* theories conforming to certain constructor-theoretic conditions. I prove that the unpredictability of measurement outcomes (to which I give an exact meaning via constructor theory), necessarily arises in superinformation theories. Then I explain how the appearance of stochasticity in (finitely many) repeated measurements can arise under superinformation theories. And I establish sufficient *conditions* for a superinformation theory to inform decisions (made under it) *as if* it were probabilistic, via a Deutsch–Wallace-type argument – thus defining a class of *decision-supporting* superinformation theories. This broadens the domain of applicability of that argument to cover constructor-theory compliant theories. In addition, in this version some of the argument's assumptions, previously construed as merely decision-theoretic, follow from *physical properties* expressed by constructor-theoretic principles.

---

[1] address for correspondence: chiara.marletto@gmail.com

## 1. Introduction

Quantum theory without the Born Rule (hereinafter: *unitary quantum theory*) is a *deterministic* theory (Wallace, 2012), whose viability as a universal physical theory has long been debated (Deutsch, 1985; Albert, 2010; Kent, 2010; Wallace, 2012). A contentious issue is how to reconcile its determinism with the *unpredictability* and *appearance of stochasticity* in quantum measurements (Albert, 2010; Saunders, 2010; Kent, 2010). Specifically, *two problems* emerge: (i) how unpredictability can occur in unitary quantum theory, since, absent the Born rule and 'collapse'-like processes, single measurements do not deliver single observed outcomes (see section 1.2); and (ii) how unitary quantum theory, despite its being non-probabilistic, can adequately account for the *appearance of stochasticity* observed in certain repeated measurements (see section 1.3).

Problems (i) and (ii) also arise in the recently proposed *constructor theory of information* (Deutsch & Marletto, 2015) (see sections 2 and 3). In that context unitary quantum theory is regarded as one of a class of theories, *superinformation theories,* most of them yet to be discovered, which are elegantly characterised by a simple, exact, constructor-theoretic condition. Specifically, certain physical systems permitted under such theories – called *superinformation media* – exhibit all the most distinctive properties of quantum systems. Like all theories conforming to the principles (Deutsch, 2013) of constructor theory, superinformation theories are expressed solely via statements about *possible and impossible tasks,* and are necessarily *non-probabilistic*. So a task being 'possible' in this sense means that it *could* be performed with arbitrarily high accuracy – *not* that it will happen with non-zero probability. Just as for unitary quantum theory, therefore, an explanation is required for how superinformation theories could account for *unpredictable measurement outcomes* and a*pparently stochastic processes*.

To provide this, I shall first provide an *exact criterion for unpredictability in constructor theory* (section 4); then I shall show that unpredictability necessarily arises in superinformation theories (and hence in quantum theory) as a result of the *impossibility of cloning* certain states – thereby addressing problem (i). Then, I shall generalise and improve upon an existing class of proposed solutions to problem (ii) in quantum theory – known as the *decision-theoretic approach* (Deutsch, 1999; Wallace, 2003, 2007, 2012), by recasting them in constructor theory. This will entail expressing a number of physical conditions on superinformation theories for them to support the decision-theoretic approach – thus defining a class of *decision-supporting superinformation theories*, which include unitary quantum theory (sections 5, 6, 7). As I shall outline in section 1.4, switching to constructor theory widens the domain of applicability of such approaches, to cover potential generalisations of quantum theory that may be discovered amongst *decision-supporting superinformation theories*; it also clarifies the assumptions on which such approaches are based, by revealing that most of them are *not* decision-theoretic, as previously thought (Deutsch 1999, Wallace 2012), but *physical*.

##### *1.1 The status of constructor theory*

Constructor theory is a proposed fundamental theory of physics (Deutsch, 2013), consisting of *principles* aiming to underlie other physical theories (such as laws of motion of elementary particles, etc.), called *subsidiary theories* in this context. Its *mode of explanation* requires physical laws to be expressed exclusively via statements about which physical transformations (more precisely, *tasks* – section 2) are *possible*, which are *impossible*, and *why*. This is a radical departure from the *prevailing conception of fundamental physics*, which instead expresses laws as *predictions* about what happens, given dynamical equations and boundary conditions in space-time.

Constructor theory is not just a framework (such as, e.g., resource theory, (Coecke, Fritz & Spekkens, 2014), or category theory (Abramsky & Coecke, 2008)) for reformulating existing theories: its principles are *proposed physical laws.* They express regularities among subsidiary theories, including new regularities that the prevailing conception cannot adequately capture. It thereby addresses some of those theories' unsolved problems. Constructor theory's principles supplement subsidiary theories, illuminating their underlying meaning and informing the development of successors, just as, for instance, the principle of energy conservation does.

In this work I appeal to the principles of the *constructor theory of information* (Deutsch & Marletto, 2015). They express the regularities in physical laws that are implicitly required by theories of information (e.g. Shannon's), via *exact* statements about possible and impossible tasks, thus giving a fully physical meaning to the hitherto fuzzily defined notion of information. Notions such as measurement and distinguishability, which are notoriously problematic to express in quantum theory yet constitute the foundation of the decision-theoretic approach, can be exactly expressed in constructor theory.

### *1.2 Unpredictability*

*Unpredictability* is a property that must be clearly distinguished from randomness. The distinction is difficult to pin down in quantum theory, especially unitary quantum theory, but can be naturally expressed in the more general context of constructor theory. Unpredictability occurs in quantum systems even given perfect knowledge of dynamical laws and initial conditions[2]. When a perfect measurer of a quantum observable $\hat{X}$ – say the *x*-component of the spin of a spin-½ particle – is presented with the particle in a superposition (or mixture) of eigenvectors of $\hat{X}$, say $|0\rangle$ and $|1\rangle$, it is *impossible* to predict reliably which outcome (0 or 1) will be observed. But in unitary quantum theory, a perfect measurement of $\hat{X}$ is merely a unitary transformation on the *source* $\mathbf{S}_a$ (the system to be measured) and the *target* $\mathbf{S}_b$ (the 'tape' of the measurer):

$$\left\{\begin{array}{ccc} |0\rangle_a|0\rangle_b & \xrightarrow{U} & |0\rangle_a|0\rangle_b \\ |1\rangle_a|0\rangle_b & \xrightarrow{U} & |1\rangle_a|1\rangle_b \end{array}\right\}$$

which implies, given the linearity of quantum theory

$$\left(\alpha|0\rangle_a+\beta|1\rangle_a\right)|0\rangle_b \xrightarrow{U} \left(\alpha|0\rangle_a|0\rangle_b+\beta|1\rangle_a|1\rangle_b\right)$$

---

[2] Thus it is sharply distinct from phenomena such as classical chaos.

for arbitrary complex amplitudes $\alpha,\beta$. Since no wave-function *collapse* is assumed, there is in reality no single "observed outcome" when the superposition is input. All possible outcomes occur simultaneously: in what sense are they unpredictable?

Additional explanations are needed – e.g. Everett's (1957) is that the observer differentiates into multiple instances, each observing a different outcome, whence the impossibility of predicting which one (Saunders, 2010; Wallace, 2012). Such accounts, however, can only ever be approximate in quantum theory, as they rely on emergent notions such as observed outcomes and 'universes' (Wallace, 2012). Also, unpredictability is a *counterfactual* property: it is not about what *will* happen, but what *cannot be made to* happen. So, while the *prevailing conception* struggles to accommodate it, constructor theory does so naturally. Just as the impossibility of cloning a set of non-orthogonal quantum states is an exact property (Wootters & Zurek, 1982), in this paper I shall express unpredictability *exactly* as a consequence of the impossibility of cloning certain sets of states under superinformation theories (section 4). My *exact*, *qualitative* characterisation of unpredictability distinguishes it from (apparent) randomness, which, as I shall explain, requires a *quantitative* explanation.

### *1.3 The appearance of stochasticity*

Another key finding of this paper is a sufficient set of conditions for superinformation theories to support a generalisation of the decision-theory approach to probability (Deutsch, 1999; Wallace, 2012), thereby explaining the *appearance of stochasticity*. This is the property that *repeated identical measurements* not only have different unpredictable outcomes, but are also, to all appearances, *random.* Specifically, consider the *frequencies* of each observed outcome[3] $x$ in multiple measurements of a quantum observable $\hat{X}$ on $N$ systems each prepared in a superposition or mixture $\rho$ of $\hat{X}$-eigenstates $|x\rangle$. The appearance of stochasticity is that for sufficiently large $N$ the frequencies do not differ significantly (according to some a-priori fixed statistical test) from the numbers $\mathrm{Tr}\{\rho|x\rangle\langle x|\}$ (and equality occurs in the limiting case of an *ensemble* (see section 6)).

[3] In the relative-state sense (Everett, 1957).

To account for this, the Born Rule states that the *probability* that $x$ is the outcome of any *individual* $\hat{X}$-measurement is $\mathrm{Tr}\{\rho|x\rangle\langle x|\}$, thus linking, *by fiat*, $\mathrm{Tr}\{\rho|x\rangle\langle x|\}$ with the frequencies in *finite* sequences of experiments. In unitary quantum theory no such link appears, prima facie, to exist, since all possible outcomes occur in reality. How can that theory be used to form an expectation about finite sequences of experiments, as its Born-rule-endowed counterpart can?

The *decision-theoretic approach* claims to solve that problem (Deutsch, 1999; Wallace, 2003, 2007, 2012; Greaves & Myrvold, 2010; Saunders, 2004, 2005). It models measurements as deterministic *games of chance*: $\hat{X}$ is measured on a superposition or mixture $\rho$ of $\hat{X}$-eigenstates; the reward is equal (in some currency) to the observed outcome. Thus the above problem is recast as that of how unitary quantum theory can inform decisions of a hypothetical rational player of that game, satisfying only *non-probabilistic* axioms of rationality. The *decision-theory argument* shows that the player, knowing unitary quantum theory (with no Born Rule) and the state $\rho$, reaches the same decision, in situations where the Born Rule would apply, *as if* they were informed by a stochastic theory with Born-Rule probabilities $\mathrm{Tr}\{\rho|x\rangle\langle x|\}$. This explains how $\mathrm{Tr}\{\rho|x\rangle\langle x|\}$ can inform expectations in single measurements under unitary quantum theory. One must additionally prove, from this, that unitary quantum theory is as *testable* as its Born-rule endowed counterpart (Wallace, 2012; Greaves & Myrvold, 2010; Deutsch, 2015).

Thus the decision-theoretic approach claims to explain the appearance of stochasticity in unitary quantum theory *without invoking stochastic laws*, rather as Darwin's theory of evolution explains the *appearance of design* in biological adaptations *without invoking a designer*. It has been challenged, especially in regard to testability (Kent, 2010; Dawid & Thébault, 2014), and defended in e.g. (Wallace, 2010; Greaves & Myrvold, 2010; Deutsch, 2015). Note that this work is *not* a defence of that approach; rather, it aims at clarifying and illuminating its assumptions (showing that most of them are physical), and at broadening its domain of applicability to more general theories than quantum theory. However, a possible application of this result may be in investigating the physical meaning of the decision-theory argument in the context of the Everett interpretation.

In my generalised version of the decision-theoretic approach, I shall define a game of chance under superinformation theories (section 7) and then identify a *sufficient set of conditions* for them to support decisions (under that approach) in the presence of unpredictability (section 6). These conditions define the class of *decision-supporting superinformation theories* (including unitary quantum theory). Specifically, they include conditions for superinformation theories to support the generalisation $f_x$ of the numbers $\mathrm{Tr}\{\rho|x\rangle\langle x|\}$ (section 5) corresponding to Born-rule probabilities. That is to say, my version of the *decision-theory argument* explains how the numbers $f_x$ can inform decisions of a player satisfying non-probabilistic rationality axioms under decision-supporting superinformation theories (section 7). Thus, decision-supporting superinformation theories would account for the appearance of stochasticity at least as adequately as unitary quantum theory.

#### *1.4 Summary of the main results*

Switching to constructor theory yields *three interrelated results*:

1) The *unpredictability* of measurements in superinformation theories is exactly distinguished from the *appearance of stochasticity*, and proved to follow from the constructor-theory generalisation of the quantum no-cloning theorem (section 4).

2) *Sufficient set of conditions* for superinformation theories to support the decision-theoretic argument (see sections 5, 6, 7) are provided, defining a *class of decision-supporting superinformation theories*, including unitary quantum theory. Constructor theory emancipates the argument from formalisms and concepts specific to (Everettian) quantum theory – such as 'observed outcomes' or 'relative states'.

3) Most premises of the decision-theory argument are no longer controversial decision-theoretic axioms, as in existing formulations, but follow from *physical properties* implied by exact principles of constructor theory.

In section 2 and 3 I summarise as much of constructor theory as is needed; in section 4 I present the *criterion for unpredictability*; in sections 5 and 6 I give the *condition for superinformation theories* to permit the constructor-theoretic generalisation of the numbers $f_x = \mathrm{Tr}\{\rho|x\rangle\langle x|\}$; in section 7, I present the decision-theory argument in constructor theory.

## 2. Constructor Theory

In constructor-theoretic physics the primitive notion of a 'physical system' is replaced by the slightly different notion of a *substrate* – a physical system some of whose properties can be changed by a physical transformation. Constructor theory's primitive elements are *tasks* (as defined below), which intuitively can be thought of as the specifications of physical transformations affecting substrates. Since tasks involving only individual states are rarely fundamental, more general descriptors for substrates are convenient:

**Attributes and variables.** The subsidiary theory must provide a collection of *states*, *attributes* and *variables*, for any given substrate. These are physical properties of the substrate, and can be represented in several interrelated ways. For example, a traffic light is a substrate, each of whose 8 states (of three lamps, each of which can be on or off) is labelled by a binary string $(\sigma_r,\sigma_a,\sigma_g):\sigma_i \in \{0,1\}, \forall i \in \{r,a,g\}$, where, say, $\sigma_r = 0$ indicates the red lamp is off, and $\sigma_r = 1$ that it is on. Similarly for $i = a$ (amber) and $i = g$ (green). Thus for instance the state where the red lamp is on and the others switched off is $(1,0,0)$.

An *attribute* is any property of a substrate that can be formally defined as a *set of all the states* in which the substrate has that property. So for example the attribute red of the traffic light, denoted by ***r***, is the set of all states in which the red lamp is on: $\boldsymbol{r} = \{(1,0,0),(1,1,0),(1,0,1),(1,1,1)\}$.

An *intrinsic* attribute is one that can be specified without referring to any other specific system. For example, 'having the same colour lamp on' is an intrinsic attribute of a pair of traffic lights, but 'having the same colour lamp on as the other one in the pair' is not an intrinsic attribute of either of them. In quantum theory,

'being entangled with each other' is, likewise, an intrinsic attribute of a qubit *pair*; 'having a particular density operator' is an intrinsic attribute of a qubit, while the rest of its quantum state describes entanglement with other systems and so is non-intrinsic.

A physical *variable* is defined in a slightly unfamiliar way as any *set of disjoint attributes* of the same substrate. In quantum theory, this includes not only all observables, but many other constructs, such as any set $\{\boldsymbol{x},\boldsymbol{y}\}$ where $\boldsymbol{x}$ and $\boldsymbol{y}$ are the attributes of being in distinct non-orthogonal states $|x\rangle$ and $|y\rangle$ of a quantum system. Whenever a substrate is in a state in an attribute $\boldsymbol{x} \in \boldsymbol{X}$, where $\boldsymbol{X}$ is a variable, we say that $\boldsymbol{X}$ is *sharp* (on that system), with the *value* $x$ – where the $x$ are members of the set $X$ of labels[4] of the attributes in $\boldsymbol{X}$. As a shorthand, "$\boldsymbol{X}$ is sharp in $\boldsymbol{a}$" shall mean that the attribute $\boldsymbol{a}$ is a subset of some attribute in $\boldsymbol{X}$. In the case of the traffic light, 'whether some lamp is on' is the variable $\boldsymbol{P} = \{\boldsymbol{off},\boldsymbol{on}\}$, where I have introduced the attributes $\boldsymbol{off} = \{(0,0,0)\}$ and $\boldsymbol{on}$, which contains all the states where at least one lamp is on. So, when the traffic light is, say, in the state $(1,0,0)$ where only the red lamp is on, we say that "$\boldsymbol{P}$ is sharp with value *on*". Also, we say that $\boldsymbol{P}$ is sharp in the attribute $\boldsymbol{r}$ (red, defined above), with value *on* – which means that $\boldsymbol{r} \subseteq \boldsymbol{on}$. In quantum theory, the *z*-component-of-spin variable of a spin-½ particle is the set of two attributes: that of the *z*-component of the spin being ½, and -½. That variable is sharp when the qubit is in a pure state with spin ½ or -½ in the *z*-direction, and is non-sharp otherwise.

**Tasks**. A *task* is the *abstract specification* of a *physical transformation* on a substrate, which is transformed from having some physical attribute to having another. It is expressed as a *set of ordered pairs of input/output attributes* $\boldsymbol{x}_i \rightarrow \boldsymbol{y}_i$ of the substrates. I shall represent it as[5]:

$$\mathfrak{A} = \{\boldsymbol{x}_1 \rightarrow \boldsymbol{y}_1,\ \boldsymbol{x}_2 \rightarrow \boldsymbol{y}_2, ...\}.$$

---

[4] I shall always define symbols explicitly in their contexts, but for added clarity I use the convention: Small Greek letters (γράμματα) denote states; ***small italic boldface*** denotes attributes; ***CAPITAL ITALIC BOLDFACE*** denotes variables; *small italic* denotes labels; *CAPITAL ITALIC* denotes sets of labels; **CAPITAL BOLDFACE** denotes physical systems; and capital letters with arrow above (e.g. $\vec{C}$ ) denote constructors.

The $\{\boldsymbol{x}_i\}$ are the legitimate *input attributes,* the $\{\boldsymbol{y}_i\}$ are the *output attributes*. A *constructor* for the task $\mathfrak{A}$ is defined as a physical system that would cause $\mathfrak{A}$ to occur on the substrates and *would remain unchanged in its ability to cause that again.* Schematically:

$$\text{Input attribute of substrates} \xrightarrow{\text{Constructor}} \text{Output attribute of substrates}$$

where constructor and substrates jointly are isolated. This scheme draws upon two primitive notions that must be given physical meanings by the subsidiary theories, namely: the substrates with the input attribute are *presented* to the constructor, which *delivers* the substrates with the output attribute. A constructor is *capable of performing* $\mathfrak{A}$ if, whenever presented with the substrates with a legitimate input attribute of $\mathfrak{A}$ (i.e., in *any* state in that attribute) it delivers them in *some* state in one of the corresponding output attributes, regardless of how it acts on the substrate with any other attribute. A task on the traffic light substrate is $\{\boldsymbol{on} \rightarrow \boldsymbol{off}\}$; and a constructor for it is a device that must switch off all its lamps whenever presented when *any* of the states in ***on***. In the case of the task $\{\boldsymbol{off} \rightarrow \boldsymbol{on}\}$ it is enough that, when the traffic light as a whole is switched off (in the state $(0,0,0)$), it delivers *some* state in the attribute ***on*** – say by switching on the red lamp only, delivering the state $(1,0,0)$ – *not necessarily all of them.*

**The fundamental principle.** A task $\mathfrak{T}$ is *impossible* if there is a law of physics that forbids its being carried out with arbitrary accuracy and reliability by a constructor. Otherwise, $\mathfrak{T}$ is *possible,* which I shall denote by $\mathfrak{T}^{\checkmark}$. This means that a constructor capable of performing $\mathfrak{T}$ can be physically realised with arbitrary accuracy and reliability (short of perfection). Catalysts and computers are familiar examples of *approximations* to constructors. So, '$\mathfrak{T}$ is possible' means that it can be brought about with arbitrary accuracy, but *it does not imply that it will happen*, since it does not imply that a constructor for it will ever be built and presented with the right substrate. Conversely, a prediction that $\mathfrak{T}$ will happen with some probability would not imply $\mathfrak{T}$'s possibility: that 'rolling a seven' sometimes happens when shooting dice does not imply that the task 'roll a seven under the rules of that game' can be performed with arbitrarily high accuracy.

[5] This '$\rightarrow$' notation for ordered pairs is intended to bring out the notion of *transformation* inherent in a task.

*Non-probabilistic, counterfactual properties* – i.e. about what *does not happen*, but could – are the centrepiece of constructor theory's mode of explanation, as expressed by its fundamental principle:

I. *All (other) laws of physics are expressible solely in terms of statements about which tasks are possible, which are impossible, and why*.

The radically different mode of explanation employed by this principle permits the formulation of *new laws of physics* (e.g. constructor information theory's ones). Thus constructor theory differs in motivation and content from existing operational frameworks, such as *resource theory (*Coecke, Fritz & Spekkens, 2014). The latter aims at proving theorems *following* from subsidiary theories, allowing their formal properties to be expressed in a unified resource-theoretic formalism. Constructor theory, in contrast, proposes *new principles*, not derivable from subsidiary theories, to supplement them, elucidate their physical meaning, and impose severe restrictions ruling out some of them.

As remarked, a constructor is closely related to the notion of a chemical catalyst, as recently formalised, e.g. in resource theory *(*Fritz, 2015*)*. A constructor is distinguished among generic catalyst-type objects in that it is required to be capable of performing the task reliably, repeatedly and to arbitrarily high accuracy. So it is not itself a physical object, but a manner of speaking about an infinite sequence of *possible* physical objects that would perform the task approximately.

Hence principle I requires subsidiary theories to have two crucial properties (holding in unitary quantum theory): (i) They must support a topology over the set of physical processes they apply to, which gives a meaning to a sequence of approximate constructions, *converging* to an exact performance of $\mathfrak{T}$; (ii) They must be non-probabilistic – since they must be expressed exclusively as statements about possible/impossible tasks. For instance, the Born-Rule-endowed versions of quantum theory, being probabilistic, do not obey the principle.

**Principle of Locality.** I shall denote the combination of two substrates $\mathbf{S}_1$ and $\mathbf{S}_2$ by $\mathbf{S}_1 \oplus \mathbf{S}_2$. Constructor theory requires all subsidiary theories to provide the following support for the concept of such a combination. First, $\mathbf{S}_1 \oplus \mathbf{S}_2$ is a substrate. Second, if subsidiary theories designate any task as possible which has $\mathbf{S}_1 \oplus \mathbf{S}_2$ as input substrate, they must provide a meaning for *presenting* $\mathbf{S}_1$ and $\mathbf{S}_2$ to the relevant constructor *as* the substrate $\mathbf{S}_1 \oplus \mathbf{S}_2$. Third, and most importantly, they must conform to Einstein's (1949) *principle of locality* in the form:

II. *There exists a mode of description such that the state of* $\mathbf{S}_1 \oplus \mathbf{S}_2$ *is the pair* $(\xi,\zeta)$ *of the states*[6] $\xi$ *of* $\mathbf{S}_1$ *and* $\zeta$ *of* $\mathbf{S}_2$, *and any construction undergone by* $\mathbf{S}_1$ *and not* $\mathbf{S}_2$ *can change only* $\xi$ *and not* $\zeta$.

Unitary quantum theory satisfies II, as is explicit in the Heisenberg picture (Deutsch & Hayden, 2000; Raymond-Robichaud & Brassard 2016). In that picture, the state of a quantum system is, at any one time, a minimal set of generators for the algebra of observables of that system, plus the Heisenberg state (Horsman & Vedral, 2007; Gottesman, 1999). Since the latter never changes, it can be abstracted away when specifying tasks: any residual 'non-locality' in that state (Wallace & Timpson, 2007) does not prevent quantum theory from satisfying principle II.

The *parallel composition* $\mathfrak{A} \otimes \mathfrak{B}$ of two tasks $\mathfrak{A}$ and $\mathfrak{B}$ is the task whose net effect on a substrate $\mathbf{M} \oplus \mathbf{N}$ is that of performing $\mathfrak{A}$ on $\mathbf{M}$ and $\mathfrak{B}$ on $\mathbf{N}$. When $\mathfrak{A} \otimes \mathfrak{T}$ is possible for some task $\mathfrak{T}$ on some generic, naturally occurring substrate (as defined in Deutsch & Marletto, 2015), $\mathfrak{A}$ is *possible with side-effects*, which is written $\mathfrak{A}^{\checkmark}$. ($\mathfrak{T}$ represents the side-effect.)

---

[6] In which case the same must hold for intrinsic attributes.

## 3. Constructor theory of information

I shall now summarise the principles of the *constructor theory of information* (Deutsch & Marletto, 2015). The principles express *exactly* the properties required of physical laws by theories of (classical) information, computation and communication – such as the possibility of copying – as well as the *exact* relation between what has been called informally 'quantum information' and 'classical information'.[7]

First, one defines *computation media.*[8] A *computation medium* with *computation variable* $\boldsymbol{V}$ (at least two of whose attributes have labels in a set $V$) is a substrate on which the task $\Pi(\boldsymbol{V})$ of performing a permutation $\Pi$ defined via the labels $V$

$$\Pi(\boldsymbol{V}) \doteq \bigcup_{x \in V} \{x \rightarrow \Pi(x)\}$$

is possible (with or without side effects), for all $\Pi$. $\Pi(\boldsymbol{V})$ is a *reversible computation*[9].

*Information media* are computation media on which additional tasks are possible. Specifically, a variable $\boldsymbol{X}$ is *clonable* if for some attribute $x_0$ of $\mathbf{S}$ the computation on the composite system $\mathbf{S} \oplus \mathbf{S}$

$$\bigcup_{x \in X} \left\{ \left(x, x_0\right) \rightarrow \left(x, x\right) \right\}, \tag{1}$$

[7] Thus, even though it is *not* itself an attempt to axiomatise quantum theory, it could provide physical foundations for information-based axiomatisations of quantum theory, (e.g. Clifton, Bub & Halvorson, 2003; Chiribella, D'Ariano & Perinotti, 2011), wherein 'information' is merely assumed to be a primitive, never explained).

[8] This is just a label for the physical systems with the given definition. Crucially, it entails no reliance on any a-priori notion of computation (such as Turing-computability).

[9] This is a *logically* reversible, (i.e., one-to-one) task, while the process implementing it may be physically irreversible, because of side-effects.

namely *cloning* $\boldsymbol{X}$, is possible (with or without side-effects)[10]. An *information medium* is a substrate with at least one clonable computation variable, called an *information variable* (whose attributes are called *information attributes*). For instance, a qubit is a computation medium with *any* set of two pure states, even if they are not orthogonal (Deutsch & Marletto, 2015); with a set of two *orthogonal* states it is an information medium. Information media must also obey the principles of constructor information theory, which I shall now recall.

**Interoperability.** Let $\boldsymbol{X_1}$ and $\boldsymbol{X}_2$ be variables of substrates $\mathbf{S}_1$ and $\mathbf{S}_2$ respectively, and $\boldsymbol{X_1} \times \boldsymbol{X}_2$ be the variable of the composite substrate $\mathbf{S}_1 \oplus \mathbf{S}_2$ whose attributes are labelled by the ordered pair $(x, x') \in X_1 \times X_2$, where $X_1$ and $X_2$ are the sets of labels of $\boldsymbol{X_1}$ and $\boldsymbol{X}_2$ respectively, and $\times$ denotes the Cartesian product of sets. The *interoperability principle* is elegantly expressed as a constraint on the composite system of information media (and on their information variables):

> III. *The combination of two information media with information variables* $\boldsymbol{X_1}$ *and* $\boldsymbol{X}_2$ *is an information medium with information variable* $\boldsymbol{X_1} \times \boldsymbol{X}_2$.

**Distinguishing and measuring** are expressed exactly in constructor theory as *tasks* involving information variables – without reference to any a priori notion of information. A variable $\boldsymbol{X}$ of a substrate $\mathbf{S}$ is *distinguishable* if

$$\left(\bigcup_{x \in X} \left\{ x \to \boldsymbol{i}_x \right\}\right)^{\checkmark} \qquad (2)$$

[10] The usual notion of cloning, as in the no-cloning theorem (Wootters & Zurek, 1982), is (1) with $\boldsymbol{X}$ as the set of all attributes of $\mathbf{S}$.

where $\{\boldsymbol{i}_x\}$ is an information variable (whereby $\boldsymbol{i}_x \cap \boldsymbol{i}_{x'} = \emptyset$ if $x \neq x'$ ). I write $\boldsymbol{x} \perp \boldsymbol{y}$ if $\{\boldsymbol{x}, \boldsymbol{y}\}$ is a distinguishable variable. Information variables themselves are distinguishable, by the interoperability principle III.[11]

A *variable* $\boldsymbol{X}$ is *measurable* if a special case of the distinguishing task (2) is possible (with or without side-effects) – namely, when the original *source substrate* continues to exist[12] in some attribute $\boldsymbol{y}_x$ and the result is stored in a *target substrate*:

$$\left(\bigcup_{x\in X}\left\{(\boldsymbol{x},\boldsymbol{x_0}) \rightarrow (\boldsymbol{y}_x, \text{‘}\boldsymbol{x}\text{’})\right\}\right)^{\checkmark} \tag{3}$$

where $\boldsymbol{x_0}$ is a generic, 'receptive' attribute and $\text{‘}\boldsymbol{X}\text{’} = \left\{\text{‘}\boldsymbol{x}\text{’} : x \in X\right\}$ is an *information variable* of the target substrate, which I shall call the *output variable* (which may, but need not, contain $\boldsymbol{x_0}$). When $\boldsymbol{X}$ is sharp on the *source w*ith any value *x*, the target is changed to having the information attribute $\text{‘}\boldsymbol{x}\text{’}$, meaning 《**S** had attribute $\boldsymbol{x}$》.

A *measurer of* $\boldsymbol{X}$ is any constructor capable of performing the task (3) for *some choice* of its output variable, labelling, and receptive state.[13] Thus, it is also is a measurer of other variables: For example, it measures any subset of $\boldsymbol{X}$, or any *coarsening* of $\boldsymbol{X}$ (a variable whose members are unions of attributes in $\boldsymbol{X}$). Two notable coarsenings of $\boldsymbol{X_1} \times \boldsymbol{X_2}$ are: $\boldsymbol{X_1} + \boldsymbol{X_2}$, where the attributes $(\boldsymbol{x_1}, \boldsymbol{x_2})$ are re-labelled with numbers $x_1 + x_2$ (and combined accordingly), and $\boldsymbol{X_1}\boldsymbol{X_2}$, where the attributes $(\boldsymbol{x_1}, \boldsymbol{x_2})$ are re-labelled with numbers $x_1 x_2$ (and likewise combined). I shall consider only *non-perturbing measurements*, i.e., $\boldsymbol{y}_x \subseteq \boldsymbol{x}$ in (3). Whenever the output variable is guaranteed to be sharp with a value $\text{‘}\boldsymbol{x}\text{’}$, I shall say, with a slight abuse of terminology, that the measurer of $\boldsymbol{X}$ "delivers a sharp output $\text{‘}\boldsymbol{x}\text{’}$".

---

[11]The set of any two non-orthogonal quantum states is not a distinguishable variable. In quantum theory two such states *on ensembles* are asymptotically distinguishable. This is generalised in constructor theory via the *ensemble-distinguishability principle*, which, in short, requires any two disjoint information attributes to be *ensemble-distinguishable* (Deutsch & Marletto, 2015).

[12] Quantum observables that are usually measured destructively, e.g. polarisation, qualify as measurable in this sense, provided that measuring them non-destructively is *possible* in principle.

[13] This differs from laboratory practice, where a measurer of $\boldsymbol{X}$ is assigned some convenient, fixed labelling of its output states.

**The 'bar' operation.** Given an information attribute $\boldsymbol{x}$, define the attribute $\bar{\boldsymbol{x}}$ ('$\boldsymbol{x}$-bar') as the union $\bigcup_{a:a\perp x} \boldsymbol{a}$ of all attributes that are distinguishable from $\boldsymbol{x}$. With this useful tool one can construct a *Boolean information variable*, defined as $\{\boldsymbol{x},\bar{\boldsymbol{x}}\}$ (which, as explained below, allows one to generalise quantum projectors). Also, for any variable **X**, define the attribute $\boldsymbol{u}_X \doteq \bigcup_{x\in X} \boldsymbol{x}$. The attribute $\bar{\bar{\boldsymbol{u}}}_X$ is the constructor-theoretic generalisation of the quantum notion of the subspace spanned by a set of states. For example, consider an information variable $\boldsymbol{X}=\{\boldsymbol{0},\boldsymbol{1}\}$ where $\boldsymbol{0}$ and $\boldsymbol{1}$ are the attributes of being in particular eigenstates of a non-degenerate quantum observable $\hat{X}$ (which also has other eigenstates). Then, $\bar{\bar{\boldsymbol{u}}}_X$ is the attribute of being in any of the possible superpositions and mixtures (prepared by any possible preparation[14]) of those two eigenstates of $\hat{X}$.

**Consistency of measurement.** In quantum theory repeated measurements of physical properties are *consistent* in the following sense. Consider the variable $\boldsymbol{X}=\{\boldsymbol{0},\boldsymbol{1}\}$ defined above. Let **2** be the attribute of being in a particular eigenstate of $\hat{X}$ orthogonal to both $\boldsymbol{0}$ and $\boldsymbol{1}$. All measurers of the variable $\boldsymbol{Z}=\{\boldsymbol{u}_X,\boldsymbol{2}\}$ are then also measurers of the variable $\boldsymbol{Z}'=\{\bar{\bar{\boldsymbol{u}}}_X,\boldsymbol{2}\}$, so that all measurers of the former, when given *any* attribute $\boldsymbol{a}\subseteq\bar{\bar{\boldsymbol{u}}}_X$, will give the *same* sharp output '$\boldsymbol{u}_X$'. The *principle of consistency of measurement* requires all subsidiary theories to have this property:

> IV. *Whenever a measurer of a variable* **Z** *would deliver a sharp output when presented with an attribute* $\boldsymbol{a}\subseteq\bar{\bar{\boldsymbol{u}}}_Z$, *all other measurers of* **Z** *would too.*

It follows (Deutsch & Marletto 2015) that they would all deliver the *same* sharp output.

**Observables.** Since (from the definition of 'bar') $\bar{\bar{\bar{\boldsymbol{x}}}}\equiv\bar{\boldsymbol{x}}$, attributes $\boldsymbol{x}$ with $\boldsymbol{x}=\bar{\bar{\boldsymbol{x}}}$ have a useful property, whence the following constructor-theoretic generalisation of quantum information observables: An *information observable* **X** *is an information*

[14] In any local formalism for quantum theory, such as the Heisenberg picture, there are many states, differently prepared, with different local descriptors, that have the same density matrix.

*variable such that whenever a measurer of* $\boldsymbol{X}$ *delivers a sharp output* $\boldsymbol{'x'}$ *the input substrate* really has the attribute $\boldsymbol{x}$.[15] A necessary and sufficient condition for a variable to be an observable is that $\boldsymbol{x} = \bar{\bar{\boldsymbol{x}}}$ for all its attributes $\boldsymbol{x}$ (Deutsch & Marletto, 2015). For example, the above-defined variable $\boldsymbol{Z} = \{\boldsymbol{u}_X, \boldsymbol{2}\}$ is not an observable (a $Z$-measurer delivers a sharp output $'\boldsymbol{u}_X'$ even when presented with a state $\xi \in \bar{\bar{\boldsymbol{u}}}_X \setminus \boldsymbol{u}_X$, where '\' denotes set exclusion), but $\{\bar{\bar{\boldsymbol{u}}}_X, \boldsymbol{2}\}$ is.

**Superinformation media.** A *superinformation medium* **S** is an information medium with at least two information observables, $\boldsymbol{X}$ and $\boldsymbol{Y}$, that contain *only mutually disjoint attributes* and *whose union is not an information observable*. $\boldsymbol{Y}$ and $\boldsymbol{X}$ are called *complementary observables*. For example, in quantum physics any set of two orthogonal qubit states constitutes an information observable, but no union of two or more such sets does: its members are not all distinguishable. *Superinformation theories* are subsidiary theories obeying constructor theory and permitting superinformation media.

From that simple property it follows that superinformation media exhibit all the most distinctive properties of quantum systems (Deutsch & Marletto, 2015). In particular, the attributes $\boldsymbol{y}$ in $\boldsymbol{Y}$ are the constructor-theoretic generalisations of what in quantum theory is called "being in a superposition or mixture" of states in the complementary observable $\boldsymbol{X}$.

**Generalised mixtures.** Consider an attribute $\boldsymbol{y} \in \boldsymbol{Y}$ and define the observable $\boldsymbol{X}_y \doteq \{\boldsymbol{x} \in \boldsymbol{X} : \boldsymbol{x} \not\perp \boldsymbol{y}\}$. (In quantum theory $\boldsymbol{X}$ could be the photon-number observable in some cavity, $|1\rangle\langle 1| + 2|2\rangle\langle 2| + \ldots$, and $\boldsymbol{y}$ the attribute of being in some superposition or mixture of some of its eigenstates, e.g. $\frac{1}{\sqrt{2}}(|0\rangle + |1\rangle)$. In that case $\boldsymbol{X}_y$ would contain two attributes, namely those of being in the states $|0\rangle\langle 0|$ and $|1\rangle\langle 1|$ respectively.) One proves (Deutsch & Marletto, 2015) that:

---

[15] "Delivering a sharp output *'x'*" means that, according to the subsidiary theory, the output variable **'X'** will be, objectively, sharp with value *'x'*; *not* (say) that the outcome *'x'* is 'observed in some universe'.

1) $\boldsymbol{X}$ is non-sharp in $\boldsymbol{y}$ since $\boldsymbol{x} \cap \boldsymbol{y} = \emptyset, \forall x \in X$ (where '$\emptyset$' denotes the empty set), and $\boldsymbol{X_y}$ contains at least two attributes.

2) Some coarsenings of $\boldsymbol{X}$ are sharp in $\boldsymbol{y}$, just as in quantum theory – where the state $\frac{1}{\sqrt{2}}(|0\rangle + |1\rangle)$ is in the +1-eigenspace of the projector $|0\rangle\langle 0| + |1\rangle\langle 1|$. The observable $\{\bar{\bar{\boldsymbol{u}}}_{X_y}, \bar{\boldsymbol{u}}_{X_y}\}$, $\boldsymbol{u}_{X_y} = \bigcup_{x \in X_y} \boldsymbol{x}$, is the constructor-theoretic generalisation of such a projector, and it is shown to be sharp in $\boldsymbol{y}$, with value $\bar{\bar{\boldsymbol{u}}}_{X_y}$. Like in quantum theory, any measurer of $\boldsymbol{X}$ presented with $\boldsymbol{y}$, followed by a computation whose output is «whether the outcome was one of the '$\boldsymbol{x}$' with $\boldsymbol{x} \not\perp \boldsymbol{y}$», will provide a sharp output '$\bar{\bar{\boldsymbol{u}}}_{X_y}$', corresponding to «***yes***». (Here and throughout, I adopt the convention that a 'quoted' attribute is the one that would be delivered by a measurement of the un-quoted one, with suitable labelling – and likewise for variables.) In summary:

| ***Quantum Theory*** <br> $\lvert y\rangle$ is an eigenstate of an observable $\hat{Y}$ with $[\hat{X},\hat{Y}] \neq 0$ | | ***Constructor Theory*** <br> $\boldsymbol{y} \in \boldsymbol{Y}; \boldsymbol{Y}$ is complementary to $\boldsymbol{X}$ |
|---|---|---|
| $\forall x \in X, \langle x \vert y\rangle \neq 1$ <br> $X_y \doteq \{x \in X : \langle x \vert y\rangle \neq 0\}$ has at least two elements | $\Leftrightarrow$ | $\boldsymbol{y} \cap \boldsymbol{x} = \emptyset$ <br> $\boldsymbol{X}_y = \{\boldsymbol{x} \in \boldsymbol{X} : \boldsymbol{x} \not\perp \boldsymbol{y}\}$ has at least two elements |
| $\hat{P} \doteq \sum_{x \in X_y} \lvert x\rangle\langle x\rvert$ | $\Leftrightarrow$ | $\boldsymbol{P} \doteq \{\bar{\bar{\boldsymbol{u}}}_{X_y}, \bar{\boldsymbol{u}}_{X_y}\}$ |
| $\mathrm{Tr}(\hat{P}\lvert y\rangle\langle y\rvert) = 1$ | $\Leftrightarrow$ | $\boldsymbol{y} \subseteq \bar{\bar{\boldsymbol{u}}}_{X_y}$ |

For any observable $\boldsymbol{H} = \{\boldsymbol{h}_1, \ldots \boldsymbol{h}_n\}$, I call an information variable $z$ a *generalised mixture of (the attributes in)* $\boldsymbol{H}$ if *either* $\boldsymbol{z}$ is in $\boldsymbol{H}$ (then it is a trivial mixture) *or* $(\forall \boldsymbol{h}_i \in \boldsymbol{H})(\boldsymbol{z} \cap \boldsymbol{h}_i = \emptyset \ \& \ \boldsymbol{z} \not\perp \boldsymbol{h}_i)$ and $\{\bar{\bar{\boldsymbol{u}}}_H, \bar{\boldsymbol{u}}_H\}$ is sharp in $\boldsymbol{z}$ with value $\bar{\bar{\boldsymbol{u}}}_H$. (In quantum theory, the $\boldsymbol{h}_i$ could be attributes of being eigenstates of some non-degenerate quantum observable, and a generalised mixture of those attributes would be a quantum superposition or mixture of those eigenstates.)

3) Let $(\boldsymbol{a}_y, \boldsymbol{b}_y)$ be the attribute[16] delivered by an $X$-measurer (with substrates $\mathbf{S}_a \oplus \mathbf{S}_b$), when presented with $\boldsymbol{y}$ (figure 1). (In quantum theory, $(\boldsymbol{a}_y, \boldsymbol{b}_y)$ could be an entangled

---

[16] It is a pair of attributes by locality: in quantum theory, for a fixed Heisenberg state, that is a pair of sets of local descriptors, generators of the algebra of observables for $\mathbf{S}_a$ and $\mathbf{S}_b$ (Gottesman, 1999; Horsman & Vedral, 2010).

state resulting from a measurement of $X$, where $y$ was the state $\frac{1}{\sqrt{2}}(|0\rangle+|1\rangle)$.) One can show that each of the local descriptors $a_y$ and $b_y$ has the same properties 1) and 2) as $y$ does, as follows:

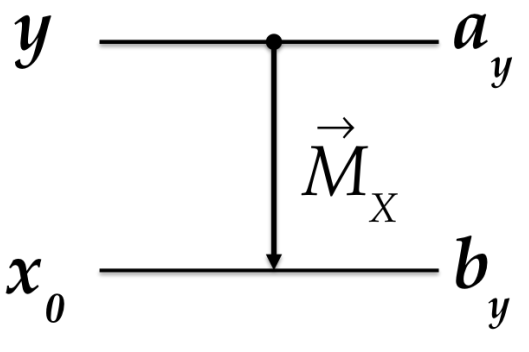

**Figure 1**

Let $'X'_y \doteq \{'x' \in 'X' : x \in X_y\}$ (i) $'X'_y$ is not sharp in $b_y$. (If it were, with value *'x'*, that would imply, via the property of observables, that $y \subseteq x$, contrary to the defining property that $x \cap y = \emptyset$.) (ii) Also, $b_y \not\perp 'x'$, $\forall 'x' \in 'X'_y$ (for if $b_y \perp 'x'$, then $y$ could be distinguished from $x$, contrary to assumption). (iii) $b_y \subseteq \bar{\bar{u}}_{'X'_y}$. For a measurer of $\{\bar{\bar{u}}_{'X'_y}, \bar{u}_{'X'_y}\}$ applied to the target substrate of an $X$-measurer is also a measurer of $\{\bar{\bar{u}}_{X_y}, \bar{u}_{X_y}\}$; hence, when presented with $y$, it must deliver a sharp output $'\bar{\bar{u}}_{'X'_y}'$. By the property of observables, $\{\bar{\bar{u}}_{'X'_y}, \bar{u}_{'X'_y}\}$ must be sharp in $b_y$, with value $\bar{\bar{u}}_{'X'_y}$. By the same argument, one shows that $X_y$ is not sharp in $a_y$, i.e. $a_y \cap x = \emptyset$; that $a_y \not\perp x$, $\forall x \in X_y$; and that $\{\bar{\bar{u}}_{X_y}, \bar{u}_{X_y}\}$ is sharp in $a_y$, with value $\bar{\bar{u}}_{X_y}$.

**Intrinsic parts of attributes.** The attributes $a_y$ and $b_y$ are not intrinsic, for each depends on the history of interactions with other systems. (In quantum theory, $\mathbf{S}_a$ and $\mathbf{S}_b$ are entangled.) However, because of the principle of locality, given an information observable $X$, one can define the *X-intrinsic part* $[a_y]_X$ of the attribute $a_y$ as follows. Consider the attribute $(a_y, b_y)$ prepared by measuring $X$ on system $\mathbf{S}_a$ using *some* particular substrate $\mathbf{S}_b$ as the target substrate. In each such preparation, $\mathbf{S}_a$ will have the same intrinsic attribute $[a_y]_X$, which I shall call the *X-intrinsic part* of $a_y$, which is therefore *the union of all the attributes preparable in that way*. The same construction defines the '$X$'-intrinsic part $[b_y]_{'X'}$ of $b_y$.

It follows, from the corresponding property of $a_y$: That $\{\bar{\bar{u}}_{X_y}, \bar{u}_{X_y}\}$ is sharp in $[a_y]_X$ with value $\bar{\bar{u}}_{X_y}$; that $[a_y]_X \cap x = \phi$; That $[a_y]_X \not\perp x$, $\forall x \in X_y$. Similarly for the 'quoted' variables and attributes. In quantum theory, $[a_y]_X$ and $[b_y]_{'X'}$ are attributes of having the reduced density matrices on $\mathbf{S}_a$ and $\mathbf{S}_b$. Unlike in (Zurek, 2005), they are *not given any probabilistic interpretation*, but only used as local descriptors of locally accessible information (defined deterministically in constructor theory (Deutsch & Marletto, 2015)).

**Successive measurements.** In unitary quantum theory the consistency of measurement (see above) is the feature that when successive measurers of $\hat{X}$ are applied to the same source initially in the state $(\alpha|0\rangle+\beta|1\rangle)$, with two systems $\mathbf{S}_b$ and $\mathbf{S}_{b'}$ as targets:

$$(\alpha|0\rangle_a+\beta|1\rangle_a)|0\rangle_b|0\rangle_{b'} \to \alpha|0\rangle_a\underbrace{|0\rangle_b|0\rangle_{b'}}+\beta|1\rangle_a\underbrace{|1\rangle_b|1\rangle_{b'}}$$

the projector for «whether the two target substrates hold the same value» is sharp with value 1. In constructor theory the generalisation of that property is required to hold. Define a useful device, the ***X**-comparer*[17] $\vec{C}_X$. It is a constructor for the task of *comparing* two instances of a substrate in regard to an observable $\boldsymbol{X}$ defined on each:

$$\bigcup_{x,x'\in X}\{(x,x',x_0)\to(x,x',c_{x,x'})\} \qquad (4)$$

where $c_{x,x'}=$ «***yes***» if $\boldsymbol{x=x'}$ (i.e. if the first two substrates (*sources*) hold attributes with the same label) and $c_{x,x'}=$ «***no***» otherwise. $\{$«***yes***»,«***no***»$\}$ is an information observable of a third substrate (*target*). In quantum theory if $\hat{X}$ has eigenstates $\{|x\rangle\}$, $\vec{C}_X$ is realised by a unitary that delivers «***yes***» (respectively «***no***») whenever the state of the sources is in $\underset{x\in X}{\text{Span}}\{|x\rangle|x\rangle\}$ or a mixture thereof (respectively $\underset{x',x\in X:x\neq x'}{\text{Span}}\{|x\rangle|x'\rangle\}$). The equivalent holds under constructor theory, by the principle of consistency of measurement IV: $\vec{C}_X$ delivers the output «***yes***» if and only if the sources hold an attribute in $\overline{\overline{\bigcup_{x\in X}\{(x,x)\}}}$. Thus, in particular, if one of the sources has the attribute $\boldsymbol{x}$ then $\vec{C}_X$ is a measurer of $\{\overline{\overline{x}},\overline{x}\}$, i.e. of whether the other source also has the attribute $\boldsymbol{x}$ (a property used in section 4).

[17] $X$ is non-bold (denoting a set of labels) in order to stress that its task is not invariant under re-labelling.

The fact that a quantum $\vec{C}_X$ would deliver a sharp «***yes***» when presented with the target substrates of successive measurements of an observable on the same source, is what makes 'relative states' and 'universes' meaningful in Everettian quantum theory, because it makes the notion of 'observed outcome in a universe' meaningful even when the input variable $\boldsymbol{X}$ of the measurer is not sharp. The same holds in superinformation theories (figure 2).

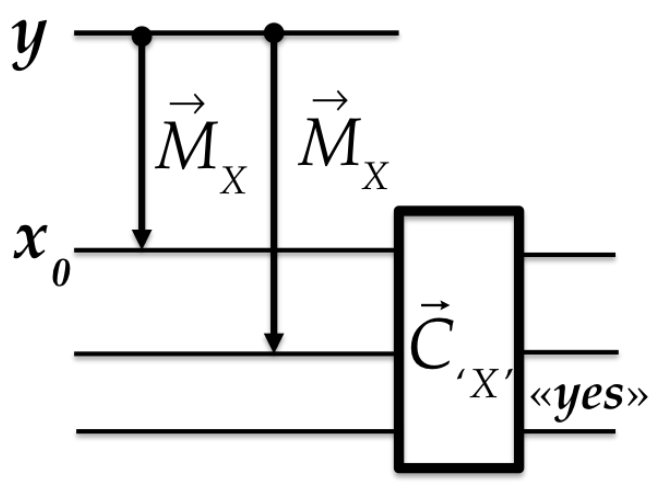

**Figure 2** Consistency of repeated measurements.

**4. Unpredictability in superinformation media**

Having presented the necessary parts of constructor information theory, I can now discuss unpredictability. I shall define it exactly in constructor theory, and show how it arises in superinformation media.

**$X$-predictor.** An *X-predictor* for the output of an $X$-measurer whose input attribute $z$ is drawn from some variable $\boldsymbol{Z}$ (in short: '*X-predictor for* $\boldsymbol{Z}$'), is a constructor for the task:

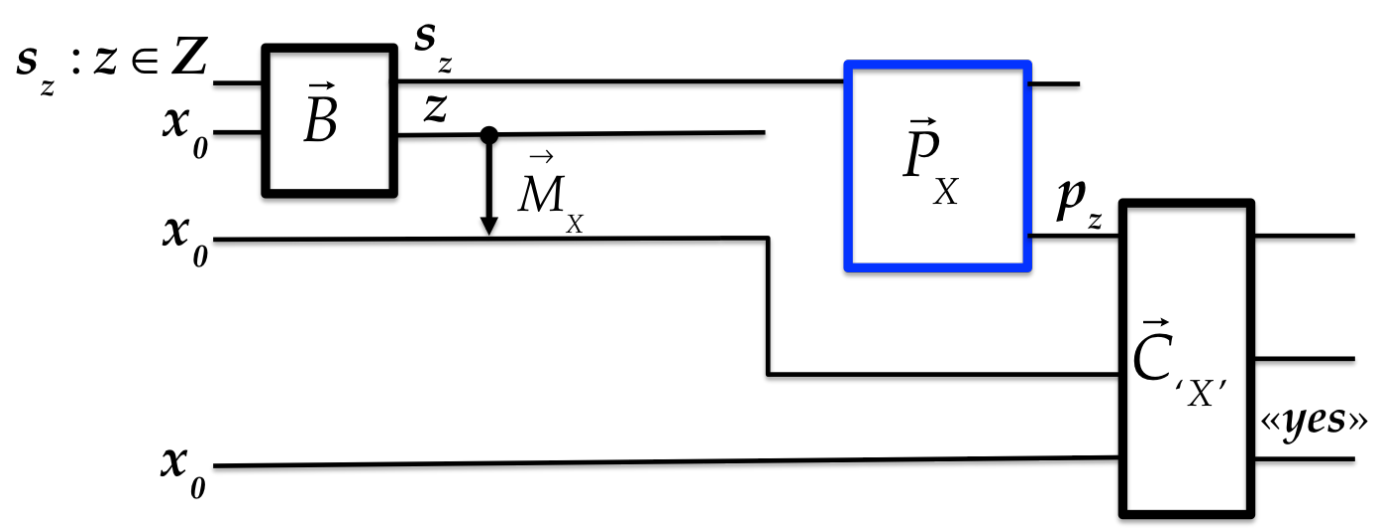

**Figure 3** The scheme defining an $X$-predictor $\vec{P}_X$

$$\bigcup_{z\in Z}\{(s_z,x_0)\to(s_z,p_z)\}$$

where $\boldsymbol{P}=\{p_z\}$ is an information observable whose attributes $p_z$ – each representing the *prediction* «the outcome of the $X$-measurer will be **'x'** given the attribute $z$ as input» – are required to satisfy the network of constructions in figure 3. $\vec{B}$ first *prepares* $\mathbf{S}_a$ with the information attribute $z\in\boldsymbol{Z}$ specified by some information attribute $s_z$; then the *X-measurer* $\vec{M}_X$ is applied to $\mathbf{S}_a$; and then its target $\mathbf{S}_b$ and the output of the predictor, $p_z$, are presented to an '$X$'-comparer $\vec{C}_{'X'}$. If that delivers a sharp «***yes***», the prediction $p_z$ is confirmed. If it would be confirmed for all $z\in\boldsymbol{Z}$, then $\vec{P}_X$ is an $X$-predictor for $\boldsymbol{Z}$.

The exact *definition of unpredictability* is then:

A *substrate exhibits unpredictability if, for some observable* **X**, *there is a variable* **Z** *such that an X-predictor for* **Z** *is impossible*

Hence unpredictability is the *impossibility* of an *X*-predictor for a variable **Z**. Note the similarity to 'no-cloning', i.e., the *impossibility* of a constructor for the cloning task (2) on the variable **Z**.

**No-cloning implies unpredictability.** Indeed, I shall now show that superinformation theories (and thus unitary quantum theory) exhibit unpredictability as a consequence of the impossibility of cloning certain sets of attributes.

Consider two complementary observables $\boldsymbol{X}$ and $\boldsymbol{Y}$ of a superinformation medium and define the variable $\boldsymbol{Z} = \boldsymbol{X}_y \cup \{\boldsymbol{y}\}$. I show *that there cannot be an X-predictor for* **Z**. For suppose there were. The predictor's output information observable $\boldsymbol{P}$ would have to include the observable $'\boldsymbol{X}'_y$. For, if $\boldsymbol{z}$=$\boldsymbol{x}$ for some $\boldsymbol{x} \in \boldsymbol{X}_y$, $'\boldsymbol{X}'_y$ has to be sharp on the target $\mathbf{S}_b$ of the measurer with value $'\boldsymbol{x}'$; so the $'\boldsymbol{X}'$-comparer yielding a sharp «*yes*» would require $\boldsymbol{p}_x = '\boldsymbol{x}'$ (as explained in section 3, $C_{'X'}$ is a measurer of $\left\{\overline{\overline{'x'}}, \overline{'x'}\right\}$ when $'\boldsymbol{X}'$ is sharp on one of its sources with value $'\boldsymbol{x}'$).

When $\boldsymbol{z}$=$\boldsymbol{y}$, the *X*-predictor's output attribute $\boldsymbol{p}_y$ must still cause the *X*-comparer to output the sharp outcome «*yes*»; also, $\boldsymbol{P} = '\boldsymbol{X}'_y \bigcup \{\boldsymbol{p}_y\}$ is required to be an information variable: hence either $\boldsymbol{p}_y = '\boldsymbol{x}'$ for some $'\boldsymbol{x}' \in '\boldsymbol{X}'_y$; or $'\boldsymbol{x}' \in \overline{\boldsymbol{u}}_{'X'_y}$. In the former case, again by considering $C_{'X'}$ as a measurer of $\left\{\overline{\overline{'x'}}, \overline{'x'}\right\}$, $'\boldsymbol{X}'_y$ would have to be sharp on the target $\mathbf{S}_b$ of the *X*-measurer, with the value $'\mathbf{x}'$; whence $\boldsymbol{y} \subseteq \boldsymbol{x}$, contrary to definition of superinformation. In the latter, since $\boldsymbol{y} \subset \overline{\overline{\boldsymbol{u}}}_{X_y}$, $\mathbf{S}_b$ would have the attribute $\overline{\overline{\boldsymbol{u}}}_{X'_y}$ (section 3) so that the $'X'$-comparer would have to output a sharp «*no*». This again contradicts the assumptions. So, there cannot exist an *X*-predictor for **Z**, just as there cannot be a cloner for **Z**, because **Z** is not an information variable.

Thus, unpredictability is predicted by the superinformation theory's deterministic[18] laws. Its *physical explanation* is given by the subsidiary theory. In Everettian quantum theory it is that there are different 'observed outcomes' across the multiverse. But constructor theory has emancipated unpredictability from 'observers', 'relative states' and 'universes', stating it as a *qualitative* information-theoretic property, just as no-cloning is.

**5. X-indistinguishability equivalence classes**

Quantum systems exhibit the *appearance of stochasticity*, which is more than mere unpredictability. Consider a quantum observable $\hat{X}$ of a *d*-dimensional system **S**, with eigenstates $|x\rangle$ and eigenvalues $x$. *Successive measurements* of $\hat{X}$ on *N* instances of **S**, each identically prepared in a superposition or mixture $\rho$ of $\hat{X}$-eigenstates, display the following *convergence property*: i) For large *N*, the fraction of replicas *delivering the observed outcome*[19] *'x'* when $\hat{X}$ is measured can be expected not to differ significantly (according to some a-priori fixed statistical test) from the number $\mathrm{Tr}\{\rho|x\rangle\langle x|\}$; ii) in an *ensemble* (infinite collection) of such replicas, each prepared in state $\rho$ (a "$\rho$-ensemble", for brevity), the fraction of instances that would give rise to an observed outcome *'x' equals* $\mathrm{Tr}\{\rho|x\rangle\langle x|\}$ (DeWitt, 1970).

But what justifies the expectation in i)? A frequentist approach to probability would simply *postulate* that the number $\mathrm{Tr}\{\rho|x\rangle\langle x|\}$ from ii) *is* the 'probability' of the outcome $x$ when $\hat{X}$ is measured on $\rho$ – which would imply (via ad-hoc methodological rules, see e.g. (Papineau, 2006; and Deutsch, 2015)) that $\mathrm{Tr}\{\rho|x\rangle\langle x|\}$ could inform decisions about finitely many measurements. In contrast, in unitary quantum theory, it is the decision-theoretic approach that establishes the same conclusion – *with no ad-hoc probabilistic assumption*. Absent that argument, the numbers $\mathrm{Tr}\{\rho|x\rangle\langle x|\}$ are just *labels of equivalence classes* within the set of superpositions and mixtures of the states $\{|x\rangle\}$. Each class, labelled by the $d$-tuple $[f_x]_{x\in X}$, $0 \le f_x \le 1$, $\sum_{x\in X} f_x = 1$, is the set of all states ρ with $\mathrm{Tr}\{\rho|x\rangle\langle x|\} = f_x$. For instance, $c_0|0\rangle + e^{i\phi}c_1|1\rangle$ belongs to the class labelled by $\left[|c_0|^2, |c_1|^2\right]$.

[18]This is so even if the initial conditions are specified with perfect accuracy in contrast with classical 'chaos'.

[19] Constructor theory's consistency of successive measurements (section 3), i.e. relative states in quantum theory, make the concept of observed outcome accurate for some purposes despite the lack of a single observed outcome.

I shall now give sufficient conditions on superinformation theories for a generalisation of those equivalence classes, which I shall call '*X-indistinguishability classes*', to exist on the set of all generalised mixtures of attributes of a given observable $\boldsymbol{X}$. One of the conditions for a superinformation theory to support the decision-theoretic argument will be that they allow for such classes (section 6).

In quantum theory, the equivalence classes are labelled by the d-tuple $[f_x]_{x\in X}$, where $f_x = \mathrm{Tr}\{\rho|x\rangle\langle x|\}$. Since the 'trace' operator need not be available in superinformation theories, to the end of constructing such equivalence classes I shall deploy a construction on *fictitious ensembles*. This is a novel mathematical construction, where properties of *ensemble* will be used to define *properties of single systems* without, of course, any probabilistic or frequentist interpretation.

At this stage, the $f_x$ *are only labels of equivalence classes.* Additional conditions will therefore be needed for the $f_x$ to inform decisions in the way that probabilities are *assumed* to do in stochastic theories (including traditional quantum theory via the Born Rule). I shall give these in section 7 via the decision-theory argument. No conclusion about decisions could possibly follow merely from what the results of measurements on an *infinite* ensemble would be, which is what my formal definition of the $f_x$ is about.

**X-indistinguishability classes.** I denote by $\mathbf{S}^{(N)}$ a substrate $\overbrace{\mathbf{S}\oplus\mathbf{S}\oplus\ldots\mathbf{S}}^{N \text{ instances}}$, consisting of $N$ replicas of a substrate $\mathbf{S}$. Let us fix an observable $\boldsymbol{X}$ of $\mathbf{S}$, whose attributes I suppose with no relevant loss of generality to be labelled by integers: $\boldsymbol{X} = \{x : x \in X\}$, where $X = \{0,1,....,d-1\}$. Let $X^{(N)} \doteq \{(x_1,x_2,x_3,...,x_N) : x_i \in X\}$ be the set of strings of length $N$ whose digits can take values in $X$, each denoted by $\underline{s} \doteq (s_1,s_2,s_3,...,s_N) : s_i \in X$. $\boldsymbol{X}^{(N)} = \{\underline{s} : \underline{s} \in X^{(N)}\}$ is an observable of $\mathbf{S}^{(N)}$. In quantum theory, supposing that $\hat{X}$ is an observable of a $d$-dimensional system $\mathbf{S}$, $\boldsymbol{X}^{(N)}$ might be $\hat{X}^{(N)} = \hat{X}_1 + d\hat{X}_2 + d^2\hat{X}_3 + ... + d^{N-1}\hat{X}_N$, whose non-degenerate eigenstates are the strings of length $N$: $|\underline{s}\rangle \doteq |s_1\rangle|s_2\rangle...|s_N\rangle | s_i \in X$.

Fix an $N$. For any attribute $\boldsymbol{x}$ in $\boldsymbol{X}$, I define a constructor $\vec{D}_{\boldsymbol{x}}^{(N)}$ for the task of counting the number of replicas that hold a sharp value $\boldsymbol{x}$ of $\boldsymbol{X}$:

$$\bigcup_{\underline{s}\in \boldsymbol{X}^{(N)}} \{\underline{s} \rightarrow f(x;\underline{s})\}$$

where the numbers $f(x;\underline{s}) \doteq \frac{1}{N}\sum_{s_i\in\underline{s}} \delta_{x,s_i}$ label the attributes of the output information variable $\boldsymbol{O}^{(N)} = \{\boldsymbol{o}^{(N)} : o \in \Phi^{(N)}\}$, with $\Phi^{(N)} = \{f_i^{(N)}\}$ denoting the set of fractions with denominator $N$: $f_i^{(N)} \doteq \frac{i}{N}$. Thus, whenever presented with the substrate $\mathbf{S}^{(N)}$ on which $\boldsymbol{X}^{(N)}$ is sharp with value $\underline{s}$ (i.e., $\mathbf{S}^{(N)}$ is in a state $\xi \in \underline{s}$), $\vec{D}_{\boldsymbol{x}}^{(N)}$ outputs the number of instances of $\mathbf{S}$ on which $\boldsymbol{X}$ is sharp with value $\boldsymbol{x}$. It could be realised, for instance, by measuring the observable *'X'* on each of the $N$ substrates in $\mathbf{S}^{(N)}$, and then by adding one unit to the output substrate, initially at *0*, for each *'x'* detected (figure 4). In quantum theory it effects a unitary operation defined by:

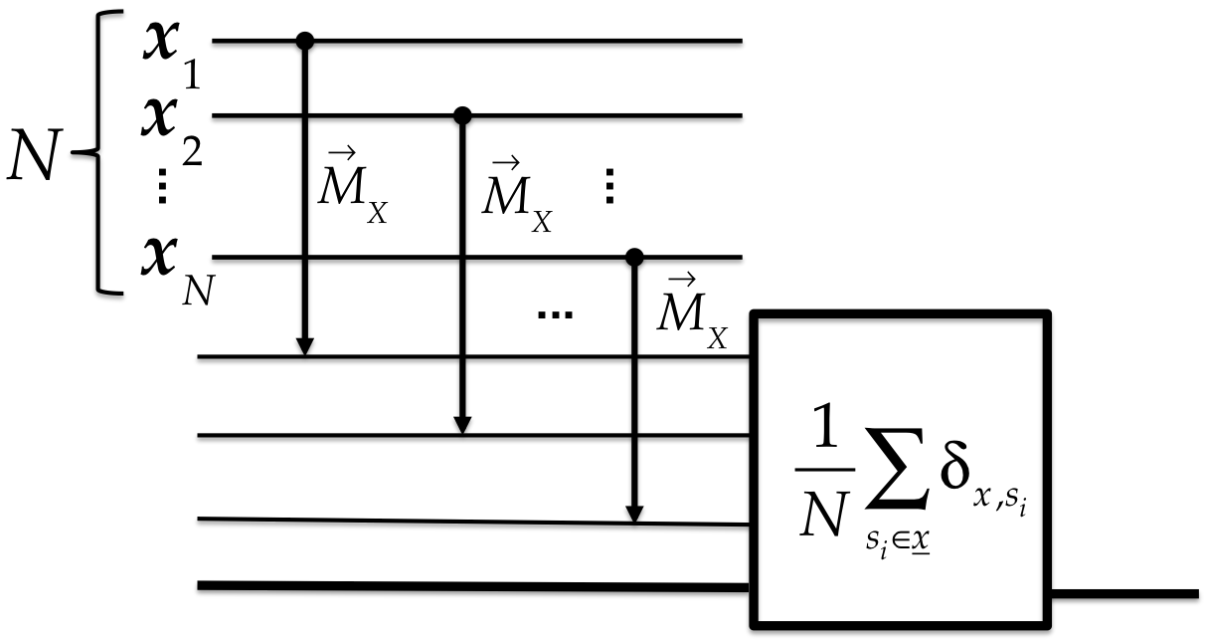

**Figure 4.** The constructor $\vec{D}_x^{(N)}$.

$$U : |\underline{s}\rangle|x_0\rangle \rightarrow |\underline{s}\rangle|f(x;\underline{s})\rangle, \forall \underline{s}$$

I shall now use $\vec{D}_{\boldsymbol{x}}^{(N)}$ to define *attributes of the substrate* $\boldsymbol{S}^{(N)}$, whose limiting case for $N \rightarrow \infty$ will be used to define the *X-indistinguishability* classes.

Consider first the observable $\boldsymbol{X}_{x,f_i^{(N)}} \doteq \{\underline{s} \in \boldsymbol{X}^{(N)} : f(x,\underline{s}) = f_i^{(N)}\}$ containing the strings $\underline{s}$ where a fraction $f_i^{(N)} \in \Phi^{(N)}$ of the replicas of $\mathbf{S}$ hold a sharp attribute $\boldsymbol{x}$. They have the property that when presented to $\vec{D}_{\boldsymbol{x}}^{(N)}$ they make the observable $\boldsymbol{O}^{(N)}$ sharp in output, with value $f_i^{(N)}$. For example, for $N$=3 and $x = 0$, $\Phi^{(3)} = \{0, \frac{1}{3}, \frac{2}{3}, 1\}$ and $\boldsymbol{X}_{0,2/3}$ contains the quantum states $\{|001\rangle, |010\rangle, |100\rangle\}$. Now define the attribute $\underline{\boldsymbol{x}}_{f_i^{(N)}}$ as the *union of all the attributes* $\boldsymbol{z}^{(N)}$ *of* $\mathbf{S}^{(N)}$ *that when presented to* $\vec{D}_{\boldsymbol{x}}^{(N)}$ *make the observable* $\boldsymbol{O}^{(N)}$ *sharp in output, with value* $f_i^{(N)}$. Next, consider the observable $\boldsymbol{F(x)}^{(N)} \doteq \left\{ \underline{\boldsymbol{x}}_{f_i^{(N)}} : f_i^{(N)} \in \Phi^{(N)} \right\}$. It follows from the consistency of measurement (section 3):

$$\underline{\boldsymbol{x}}_{f_i^{(N)}} = \bar{\bar{\boldsymbol{u}}}_{\boldsymbol{X}_{f_i^{(N)}}} .$$

Therefore, crucially, for a given $\boldsymbol{x}$, $\boldsymbol{F(x)}^{(N)}$ can be sharp even if the observable $\boldsymbol{X}^{(N)}$ is not. In the example above, for $N$=3 and $x=0$, $\underline{\boldsymbol{0}}_{2/3} \in \boldsymbol{F(0)}^{(3)}$ is the set of all superpositions and mixtures of the eigenstates of $\hat{\boldsymbol{X}}^{(3)}$ contained in $\boldsymbol{X}_{\boldsymbol{0},2/3}$, $\{|001\rangle, |010\rangle, |100\rangle\}$: $\hat{\boldsymbol{X}}^{(3)}$ is not sharp in most such mixtures and superpositions.

The observable $\boldsymbol{F(x)}^{(N)}$ is key to generalising quantum theory's convergence property, for the latter is due to the fact that there exists the limit of the sequence of attributes $\underline{\boldsymbol{x}}_{f_i^{(N)}}$ for $N \to \infty$. Let me now recall the formal expression of the *convergence property* in quantum theory (DeWitt, 1970)[20]:

Consider a state $|z\rangle = \sum_{x \in X} c_x |x\rangle$ with the property that $|z\rangle^{\otimes N} = \sum_{\underline{s}} c_{s_1} c_{s_2} \ldots c_{s_N} |\underline{s}\rangle$ is a superposition of states $|\underline{s}\rangle = |s_1\rangle |s_2\rangle \ldots |s_N\rangle \mid s_i \in X$ each having a different $f(x;\underline{s}) = f_i^{(N)}$. The convergence property is that for any positive, arbitrarily small $\varepsilon$:

$$|z\rangle^{\otimes N} \underset{N \to \infty}{\longrightarrow} \sum_{\underline{s}:\delta(\underline{s}) \le \varepsilon} c_{s_1} c_{s_2} \ldots c_{s_N} |\underline{s}\rangle$$

where $\delta(\underline{s}) = \sum_{x \in X} \left( f(x;\underline{s}) - |c_x|^2 \right)^2$. In other words, for any arbitrarily small $\varepsilon$, there exists an $N$ such that the quantum implementation of $\vec{D}_{\boldsymbol{x}}^{(N)}$, when asked whether the fraction $f(x;\underline{s})$ of observed outcomes *'x'* obtained when measuring $\hat{X}^{(N)}$ on $|z\rangle^{\otimes N}$ is within $\varepsilon$ of the value $|c_x|^2$, is in a state as close (*in the natural Hilbert space norm provided by quantum theory*) as desired to one in which it answers «***yes***». Thus the proportion of instances delivering the observed outcome *'x'* when $\hat{X}$ is measured on the ensemble state $|z\rangle^{\infty} = \lim_{N\to\infty} |z\rangle^{\otimes N}$ *is equal* to $\mathrm{Tr}\{\rho_z |x\rangle\langle x|\} = |c_x|^2$ (where $\rho_z \doteq |z\rangle\langle z|$). Therefore, all states $|z\rangle^{\otimes N}$ with the property $\mathrm{Tr}\{\rho_z |x\rangle\langle x|\} = |c_x|^2$ will be grouped by $\vec{D}_{\boldsymbol{x}}^{(N)}$ in the *same set*, as $N$ tends to infinity, which can thus be labelled by $\mathrm{Tr}\{\rho_z |x\rangle\langle x|\}$. This set is an attribute of a *single* system, containing all quantum states with the property that $\mathrm{Tr}\{\rho_z |x\rangle\langle x|\} = |c_x|^2 = f_x$. The set of all superpositions and mixtures of eigenstates of $\boldsymbol{X}$ is thus partitioned equivalence classes, labelled by the *d*-tuple $[f_x]_{x\in X}$, $0 \le f_x \le 1$, $\sum_{x\in X} f_x = 1$.

[20] My appealing to that property is *not* to conclude, using DeWitt's own words, that the "conventional statistical interpretation of quantum mechanics thus emerges from the formalism itself" (which would be circular); I shall only use it formally to construct certain attributes of the *single* substrate.

A sufficient condition on a superinformation theory for a generalisation of these '*X-indistinguishability classes*' to exist under it, on the set of all generalised mixtures of attributes of a given observable $\boldsymbol{X}$, is that it satisfy the following *requirements*:

E1) For each $\underline{\boldsymbol{x}}_{f_i^{(N)}} \in \boldsymbol{F(x)}^{(N)}$, there exists the attribute of the ensemble of replicas of **S** defined as

$$\underline{\boldsymbol{x}}_{f^{\infty}} \doteq \lim_{N\to\infty} \underline{\boldsymbol{x}}_{f_i^{(N)}},$$

where $f^{\infty} = \lim_{N\to\infty} f_i^{(N)} \in \Phi$; $\Phi$ is the limiting set of $\Phi^{(N)}$ - which must exist, and its elements, which are real numbers, must have the property $\sum_{f^{\infty}\in\Phi} f^{\infty} = 1,\ 0 \le f^{\infty} \le 1$.

The existence of the limit implies that those attributes do not intersect, i.e., the set $\boldsymbol{F(x)}^{(\infty)} = \{\underline{\boldsymbol{x}}_{f^{\infty}} : f^{\infty} \in \Phi\}$ is a (formal) variable of the ensemble – a limiting case of $\boldsymbol{F(x)}^{(N)}$, generalising its quantum analogue.

Given an attribute z of **S**, define $z^{(N)} \doteq \overbrace{(z, z, \ldots z)}^{\text{N terms}}$, (in quantum theory, this is the attribute of being in the quantum state $|z\rangle^{\otimes N}$); introduce the auxiliary variable $\boldsymbol{X}_{f^{\infty}} \doteq \left\{ z : \lim_{N\to\infty} z^{(N)} \subseteq \underline{\boldsymbol{x}}_{f^{\infty}} \right\}$ and let $\boldsymbol{x}_{f^{\infty}} \doteq \bigcup_{z\in \boldsymbol{X}_{f^{\infty}}} z$ – which, unlike its ensemble counterpart $\underline{\boldsymbol{x}}_{f^{\infty}}$, is an *attribute of a single substrate* ***S***.

E2) For any generalised mixture $\boldsymbol{z}$ of the attributes in $\boldsymbol{X}$, there exists a *d-tuple* $\left[f(z)_x\right]_{x\in X}$ with $\sum_{x\in X} f(z)_x = 1$, $0 \le f(z)_x \le 1$, such that $\boldsymbol{z} \subseteq \bigcap_{x\in X} \boldsymbol{x}_{f(z)_x}$. I shall call $\left[f(z)_x\right]_{x\in X}$ the *X-partition of unity* for the attribute $\boldsymbol{z}$. [21]

If $\boldsymbol{z}$ has such a partition of unity, it must be unique because of E1). An *X-indistinguishability equivalence class* is defined as the set of all attributes with the same *X*-partition of unity: any two attributes within that class cannot be distinguished *by measuring only the observable* $\boldsymbol{X}$ on each individual substrate, even in the limit of an infinite ensemble. In quantum theory, $\boldsymbol{x}_f$ contains all states ρ with $\mathrm{Tr}\{\rho|x\rangle\langle x|\} = f$. A superinformation theory *"admits X-partitions of unity"* (on the set of generalised mixtures of attributes in *X*) if conditions E1) and E2) are satisfied.

A key innovation of this paper is showing how the mathematical construction of an abstract infinite ensemble (culminating in the property E1)) can define structure on *individual* systems (via property E2)) – the attributes $\boldsymbol{x}_f$ and the *X*-partition of unity – *without* recourse to the frequency interpretation of probability or any other probabilistic assumption.

**X-partition of unity of the X-intrinsic part.** Consider now the attribute of being in the quantum state $|z\rangle = c_0|0\rangle + e^{i\phi}c_1|1\rangle$ whose *X*-partition of unity is $\left[|c_0|^2, |c_1|^2\right]$. In quantum theory, the reduced density matrices of the source and target substrate as delivered by an *X*-measurer acting on $|z\rangle$ still have the same partition of unity. The same holds in constructor theory. Consider the *X-intrinsic part* $[\boldsymbol{a}_z]_X$ (section 4) of the attribute $\boldsymbol{a}_z$ generated by measuring $\boldsymbol{X}$ on the attribute $\boldsymbol{z}$. By definition of $\vec{D}_{\boldsymbol{x}}^{(N)}$,

[21] The existence of $\boldsymbol{x}_{f^\infty}$ does not require there to be any corresponding attribute of the single system *for finite N*: $\boldsymbol{x}_{f^\infty}$ is constructed as a limit of a sequence of attributes *on the ensemble*. For example in quantum theory most sets $\boldsymbol{X}_{f_i^{(N)}} \doteq \{z : \boldsymbol{z}^{(N)} \subseteq \boldsymbol{x}_{f_i^{(N)}}\}$ are empty, for any *N*, except for $f_i^{(N)} = 1$, containing $\boldsymbol{x}^{(N)}$; and $f_i^{(N)} = 0$, containing all attributes $\tilde{\boldsymbol{x}}^{(N)} : \tilde{x} \in X, \tilde{x} \ne x$.

prepending a measurer of $\boldsymbol{X}$ to each of the input substrates of $\vec{D}_x^{(N)}$ will still give a $\vec{D}_x^{(N)}$, with the same labellings. Thus the construction that would classify $\boldsymbol{z}$ as being in a certain $X$-partition of unity, can be reinterpreted as providing a classification of $[\boldsymbol{a}_z]_X$, under the same labellings: the two classifications must coincide. *If $\boldsymbol{y}$ has a given X-partition of unity, the X-intrinsic part $[\boldsymbol{a}_z]_X$ of $\boldsymbol{a}_z$ must have the same one.* Likewise for the intrinsic part $[b_z]_X$ of $\boldsymbol{b}_z$ (obtained as output attribute on the *target* of the $X$-measurement applied to $\boldsymbol{y}$): the '$X$'-partition of unity of $[b_z]_X$ is numerically the same as the $X$-partition of unity of $\boldsymbol{z}$.

## 6. Conditions for decision-supporting superinformation theories

For a given observable $\boldsymbol{X}$ and a generalised mixture $\boldsymbol{z}$ of attributes in $\boldsymbol{X}$, the labels $\left[f(z)_x\right]_{x\in X}$ of the $X$-partition of unity defined in section 6 are *not* probabilities. Even though they are numbers between 0 and 1, and sum to unity, they need not satisfy other axioms of the probability calculus: for instance, in quantum interference experiments they do not obey the axiom of additivity of probabilities of mutually exclusive events (Deutsch, Ekert & Lupacchini, 2000). It is the decision-theory argument (section 7) that explains under what circumstances the numbers $\left[f(z)_x\right]_{x\in X}$ can be used to inform decisions in experiments on finitely many instances *as if* they were probabilities, without assuming them to be so. I shall now establish *sufficient conditions* on superinformation theories to support the decision-theory argument, thus characterising *decision-supporting superinformation theories*.

I shall introduce one of the conditions via the special case of quantum theory. Consider the $x$- and $y$-components of a qubit spin, $\hat{X}$ and $\hat{Y}$. There exist eigenstates of $\hat{X}$, i.e. $|x_1\rangle, |x_2\rangle$, and of $\hat{Y}$, i.e. $|y_\pm\rangle = \frac{1}{\sqrt{2}}\left[|x_1\rangle \pm |x_2\rangle\right]$ with the property that they are 'equally weighted', respectively, in the $x$- and $y$- basis – in other words, $|x_1\rangle, |x_2\rangle$ are invariant under the action of a unitary that swaps $|y_+\rangle$ with $|y_-\rangle$; and $|y_\pm\rangle$ are invariant under a unitary that swaps $|x_1\rangle$ with $|x_2\rangle$. Moreover, there exist quantum states on the composite system of two qubits, which are likewise 'equally weighted' and have the special property that:

$$\frac{1}{\sqrt{2}}\left[|x_1\rangle|x_1\rangle \pm |x_2\rangle|x_2\rangle\right] = \frac{1}{\sqrt{2}}\left[|y_+\rangle|y_+\rangle \pm |y_-\rangle|y_-\rangle\right] \quad (5)$$

I shall now require that the analogous property holds in superinformation theories. While in quantum theory it is straightforward to express this property via the powerful tools of linear superpositions, in constructor theory expressing the same conditions will require careful definition in terms of 'generalised mixtures'.

The conditions for decision-supporting information theories is that there exist two complementary information observables $\boldsymbol{X}$ and $\boldsymbol{Y}$ such that:

T1) The theory admits $X$-partitions of unity (on the information attributes of $\mathbf{S}$ that are generalised mixtures of attributes of $\boldsymbol{X}$) and $X_a + X_b$-partitions of unity (on the information attributes of the substrate $\mathbf{S}_a \oplus \mathbf{S}_b$ that are generalised mixtures of the attributes in the observable $\boldsymbol{X}_a + \boldsymbol{X}_b$).

T2) There exist observables $\tilde{\boldsymbol{X}} \doteq \{\boldsymbol{x}_1, \boldsymbol{x}_2\} \subseteq \boldsymbol{X}$, $\tilde{\boldsymbol{Y}} \doteq \{\boldsymbol{y}_+, \boldsymbol{y}_-\} \subseteq \boldsymbol{Y}$ satisfying the following *symmetry requirements*:

R1. $\boldsymbol{x}_1, \boldsymbol{x}_2$ are generalised mixtures of $\tilde{\boldsymbol{Y}}$ and $\{\boldsymbol{y}_1, \boldsymbol{y}_2\}$ are generalised mixtures of attributes of $\tilde{\boldsymbol{X}}$.

As a consequence, since $\left\{\bar{\bar{\boldsymbol{u}}}_{\tilde{X}}, \bar{\boldsymbol{u}}_{\tilde{X}}\right\}$ is sharp in both $y_+$ and $y_-$ with value $\bar{\bar{\boldsymbol{u}}}_{\tilde{X}}$, it follows that $f(y_+)_x = 0 = f(y_-)_x$ for all $x$ other than $x_1$ and $x_2$. Similarly, $f(x_1)_y = 0 = f(x_2)_y, \forall y \neq y_+, y_-$. Note also that by definition of complementary observables, $x_i \cap y_\pm = \emptyset$ – i.e., the attributes are *non-trivial* generalised mixtures.

Defining the computation $S_{a,b} \doteq \{\boldsymbol{a} \to \boldsymbol{b}, \boldsymbol{b} \to \boldsymbol{a}\}$ which swaps the *attributes* $\boldsymbol{a}, \boldsymbol{b}$ of $\mathbf{S}$:

R2. $S_{x_1,x_2}(\boldsymbol{y}_\pm) \subseteq \boldsymbol{y}_\pm$; $S_{y_+,y_-}(\boldsymbol{x}_i) \subseteq \boldsymbol{x}_i, i = 1,2$.

In quantum theory, $\boldsymbol{y}_\pm$ correspond to two distinguishable equally-weighted quantum superpositions or mixtures of the attributes in $\boldsymbol{X}_y$, such as $|y_\pm\rangle$. Similarly for $|x_1\rangle, |x_2\rangle$. The principle of consistency of measurement (section 3) implies that if an attribute $\boldsymbol{y}$ has an $X$-partition of unity with element $f(y)_x$, for any permutation $\Pi$ on $\boldsymbol{X}$, $\Pi(\boldsymbol{y})$ has $X$-partition of unity such that $f(\Pi(\boldsymbol{y}))_x = f(\boldsymbol{y})_{\Pi(x)}$. This is because presenting $\Pi(\boldsymbol{y})$ to $\vec{D}_x^{(N)}$ (defined in section 5) is equivalent to presenting $\boldsymbol{y}$ to the constructor (measurer) obtained prepending the computation $\Pi$ to $\vec{D}_x^{(N)}$ - i.e., the constructor $\vec{D}_{\Pi(x)}^{(N)}$. Therefore $S_{x_1,x_2}(\boldsymbol{y}_\pm) \subseteq \boldsymbol{y}_\pm$ implies $f(y_\pm)_{x_1} = f(y_\pm)_{x_2} = 1/2$. The same, *mutatis mutandis,* holds for the attributes $x_1, x_2$ .

Let the attributes $[\boldsymbol{a}_{y_\pm}]_X \supseteq \boldsymbol{a}_{y_\pm}$ be the $X$-intrinsic parts of the attributes $\boldsymbol{a}_{y_\pm}$, where $(\boldsymbol{a}_{y_\pm}, \boldsymbol{b}_{y_\pm})$ is the attribute of $\mathbf{S}_a \oplus \mathbf{S}_b$ prepared by measuring $\boldsymbol{X}$ on $\mathbf{S}_a$ holding the attribute $\boldsymbol{y}_\pm$, with $\mathbf{S}_b$ as a target.

R3. $[\boldsymbol{a}_{y_+}]_X = [\boldsymbol{a}_{y_-}]_X$, with $S_{x_1,x_2}\left([\boldsymbol{a}_{y_\pm}]_X\right) \subseteq [\boldsymbol{a}_{y_\pm}]_X$

This requirement is satisfied in quantum theory: a measurer of the quantum observable $\hat{X}$ acting on a substrate $\mathbf{S}_a$ in the state $|y_\pm\rangle$ generates the quantum states:

$$\frac{1}{\sqrt{2}}\left[|x_1\rangle|x_1\rangle \pm |x_2\rangle|x_2\rangle\right]$$

whose reduced density operators on $\mathbf{S}_a$ are the same, $\frac{1}{2}\left[|x_1\rangle\langle x_1| + |x_2\rangle\langle x_2|\right]$, and are still 'equally weighted', thus invariant under swap of $|x_1\rangle$ and $|x_2\rangle$.

Consider now the observables of the *composite system* $\boldsymbol{S} \oplus \boldsymbol{S}$:

$$S_x \doteq \{(\boldsymbol{x}_1, \boldsymbol{x}_1), (\boldsymbol{x}_2, \boldsymbol{x}_2)\} \subseteq X \times X$$
$$S_y \doteq \{(\boldsymbol{y}_+, \boldsymbol{y}_+), (\boldsymbol{y}_-, \boldsymbol{y}_-)\} \subseteq Y \times Y$$

R4. There exists an attribute $\boldsymbol{q}$ that is *both* a generalised mixture of attributes in $S_x$ *and* a generalised mixture of attributes in $S_y$:

$$\boldsymbol{q} \subseteq \bar{\bar{u}}_{S_x} \cap \bar{\bar{u}}_{S_y}, \quad \boldsymbol{q} \not\angle (\boldsymbol{x}_i, \boldsymbol{x}_i), \quad \boldsymbol{q} \cap (\boldsymbol{x}_i, \boldsymbol{x}_i) = \emptyset, \quad (i = 1,2)$$
$$\boldsymbol{q} \not\angle (\boldsymbol{y}_\pm, \boldsymbol{y}_\pm), \quad \boldsymbol{q} \cap (\boldsymbol{y}_\pm, \boldsymbol{y}_\pm) = \emptyset$$

with the property that:

$$S_{x_1x_1,x_2x_2}(\boldsymbol{q}) \subseteq \boldsymbol{q}, \quad S_{y_+y_+,y_-y_-}(\boldsymbol{q}) \subseteq \boldsymbol{q} \quad . \tag{6}$$

where the swap $S$ on a pair of substrates acts in parallel in each separately. In quantum theory, $\boldsymbol{q}$ is the attribute of being in the state $\frac{1}{\sqrt{2}}\left[|x_1\rangle|x_1\rangle+|x_2\rangle|x_2\rangle\right]$, because of the property (5).

These conditions define a class of *decision-supporting superinformation theories*. (See table 1 for a summary, and see also footnote 27). They are satisfied by quantum theory, as we said, via the existence of states such as those for which (5) holds.

| |
|---|
| A superinformation theory is *decision-supporting* if there exist complementary observables $\boldsymbol{X}$ and $\boldsymbol{Y}$ such that:<br><br>1) The theory admits ***X**-partitions of unity* (conditions E1), E2) in section 5) on the substrate **S** and $(X_a+X_b)$-partitions of unity on the substrate $\mathbf{S}_a\oplus\mathbf{S}_b$.<br><br>2) There exist attributes $\boldsymbol{x}_1,\boldsymbol{x}_2\in X$ that are generalised mixtures of the attributes $\{\boldsymbol{y}_+,\boldsymbol{y}_-\}$; and attributes $\boldsymbol{y}_+,\boldsymbol{y}_-\in\boldsymbol{Y}$ that are generalised mixtures of attributes $\{\boldsymbol{x}_1,\boldsymbol{x}_2\}$ with the property that:<br><br>• $S_{x_1,x_2}(\boldsymbol{y}_\pm)\subseteq\boldsymbol{y}_\pm$; $S_{y_+,y_-}(\boldsymbol{x}_i)\subseteq\boldsymbol{x}_i, i=1,2$.<br><br>• $[a_{y_+}]_X=[a_{y_-}]_X$ and $S_{x_1,x_2}([a_{y_\pm}]_X)\subseteq[a_{y_\pm}]_X$<br><br>3) There exists an attribute $\boldsymbol{q}$ that is a generalised mixture of attributes in $S_x=\{(\boldsymbol{x}_1,\boldsymbol{x}_1),(\boldsymbol{x}_2,\boldsymbol{x}_2)\}$ *and* a generalised mixture of attributes in $S_y=\{(\boldsymbol{y}_+,\boldsymbol{y}_+),(\boldsymbol{y}_-,\boldsymbol{y}_-)\}$, with the property that<br><br>$S_{x_1x_1,x_2x_2}(\boldsymbol{q})\subseteq\boldsymbol{q}, S_{y_+y_+,y_-y_-}(\boldsymbol{q})\subseteq\boldsymbol{q}$ |

*Table 1: Conditions for decision-supporting superinformation theories*

**7. Games and decisions with superinformation media**

The key step in the decision-theoretic approach in quantum theory is to model the physical processes displaying the appearance of stochasticity as *games of chance,* played with equipment obeying unitary (hence non-probabilistic) quantum theory.

In the language of the partition of unity, applied to quantum theory, the game is informally characterised as follows. An $\boldsymbol{X}$-measurer is applied with an input attribute $\boldsymbol{y}$ whose $X$-partition of unity is $\left[f(y)_{x_1},...f(y)_{x_d}\right]$. The game-device is such that if the observable $\boldsymbol{X}$ is sharp in $\boldsymbol{y}$ with value $x$ (so under quantum theory, $\boldsymbol{y}$ is an eigenstate of $\hat{X}$ with eigenvalue $x$), the reward is '$x$' (in some currency); otherwise unpredictability arises (section 4). The player may pay for the privilege of playing the game, knowing $\boldsymbol{y}$, the rules of the game, and unitary quantum theory (i.e. a full description of the physical situation). One proves that a player satisfying the same non-probabilistic axioms of rationality as in classical decision theory (Luce & Raiffa, 1957), but without assuming the Born Rule or any other rule referring to probability, would place bets *as if* he had assumed both the identification between the $f_x$ and the probabilities of outcomes given by the Born Rule, and the ad-hoc methodological rule connecting probabilities with decisions (Papineau, 2006; Deutsch, 2015).

I shall now recast the decision-theoretic approach (Deutsch, 1999; Wallace, 2012) in constructor theory. The key differences from previous versions (Deutsch 1999, Wallace 2012) are: a) those versions involved the notions of observed outcomes and relative states, which are powerful tools in Everettian quantum theory. In constructor theory, instead, none of those will be explicitly relied upon; b) Some of the axioms that were previously considered decision-theoretic follow from properties of information media and measurements, as expressed in the constructor theory of information.

### *7.1 Games of chance*

From now on, I shall assume that we are dealing with a decision-supporting superinformation theory – i.e., a superinformation theory satisfying the conditions given in section 6, with two complementary observables $\boldsymbol{X}$ and $\boldsymbol{Y}$. For simplicity, I shall assume $\boldsymbol{X}$ and $\boldsymbol{Y}$ to be real-valued, whereby $X$ and $Y$ are the set of real numbers. In this context, I shall adopt a slightly more detailed model of a game of chance, whose centrepiece is an *X-adder*. It is defined as the constructor $\vec{\Sigma}_X$ performing the following computation on an information medium $\mathbf{S}\oplus\mathbf{S}_p\oplus\mathbf{S}_p$:

$$\bigcup_{x \in X, p \in X_p} \left\{ \left( x, p, 0 \right) \rightarrow \left( x, p, x+p \right) \right\} \qquad (7)$$

An $X$-adder is a measurer of the observable $\mathbf{X}+\mathbf{X}_p$ of the substrate $\mathbf{S}\oplus\mathbf{S}_p$ *labelled* so that when the observable $\mathbf{X}$ is sharp in input with value $x$ and the observable $\mathbf{X}_p$ is sharp in input with value *0*, $\mathbf{X}_p$ is sharp on the output payoff substrate with value $x$. For any fixed $\boldsymbol{p}$, this is also a measurer (as defined in (3)) of the observable $\mathbf{X}$ on $\mathbf{S}$.[22] In quantum theory, it is realised by a unitary satisfying $U:|x\rangle|p\rangle|0\rangle \rightarrow |x\rangle|p\rangle|p+x\rangle, \forall x,p$.

An $X$-game[23] of chance $G_X(z)$ is a *construction* defined as follows:

G1) The *game substrate* $\mathbf{S}$ (e.g. a die), with *game observable* $\mathbf{X}=\left\{\boldsymbol{x}: x \in X\right\}$, is prepared with some *legitimate game attribute* $\boldsymbol{z}$, defined as any information attribute that admits an $X$-partition of unity (where $\mathbf{X}$ may be non-sharp in it) .

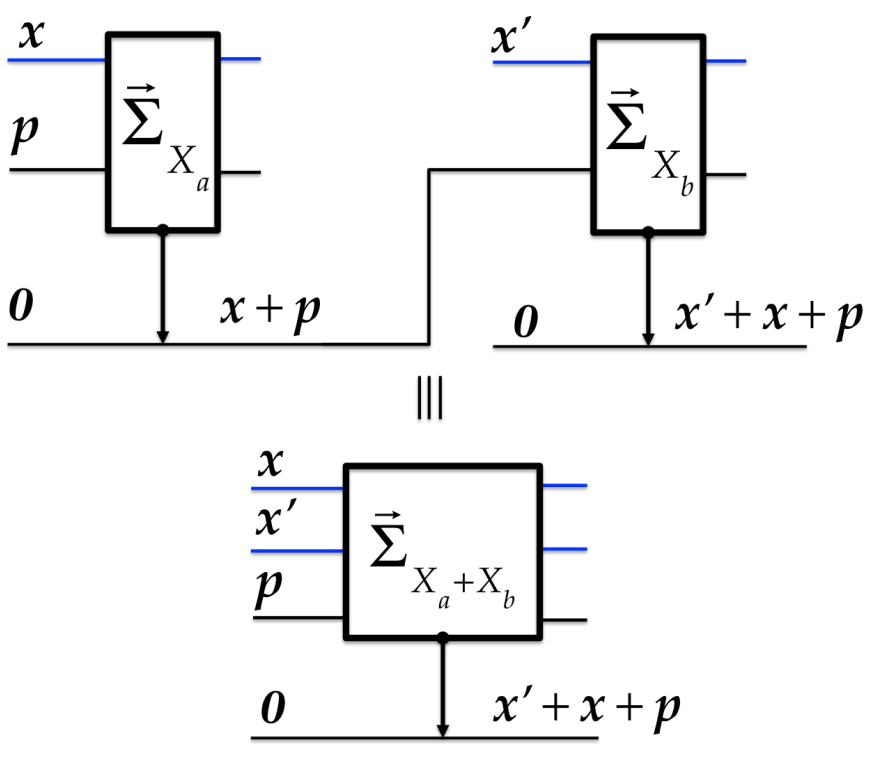

Figure 5

G2) $\mathbf{S}$ is then presented in input to an adder together with two other *payoff substrates* $\mathbf{S}_p$, representing the player's records of the winnings, with *payoff observable* $\mathbf{X}_p=\left\{\boldsymbol{x}: x \in X\right\}$. The first instance of $\mathbf{S}_p$ (the *input payoff substrate)* contains a record of the initial (pre-game) assets, in some units; the other instance (the *output payoff substrate),* initially at *0*, contains the record of the winnings at the end of the game and it is set during the game to its payoff under the action of the adder. Keeping track of both those records is an artefact of my model, superfluous in real life, but it makes it easier to analyse composite games.

---

[22] An adder is a measurer of the *same* observable $X$ for all $\boldsymbol{p}$*:* it distinguishes between any two different attributes in $X$ for any $\boldsymbol{p}$. (This is a non-trivial property: a multiplier, for instance, would not do so for $\boldsymbol{p}$=$\boldsymbol{0}$).

[23] Under constructor theory a game of this sort must be defined with respect to given labellings of the attributes. E.g. if the slots on a roulette wheel were re-labelled, a ball landing in a particular slot would cause different players to win.

G3) The *composition of two games of chance* $G_X(z)\odot G_X(z')$, with game substrates $\mathbf{S}_a$ and $\mathbf{S}_b$ is defined as the construction where the output payoff substrate of $G_X(z)$ is the input payoff substrate of $G_X(z')$ (figure 5, top). In other words, the composite game $G_X(z)\odot G_X(z')$ is an $(\boldsymbol{X}_a+\boldsymbol{X}_b)$-adder realised by measuring the observable $\boldsymbol{X}_a+\boldsymbol{X}_p$ and the observable $\boldsymbol{X}_b+\boldsymbol{X}_p$ separately. Thus, the composition of two games $G_X(z)\odot G_X(z')$ is the game $G_{X_a+X_b}((z,z'))$ with game substrate $\mathbf{S}_a\oplus\mathbf{S}_b$ (figure 5, bottom).

**The player.** I shall model the player for $G_X(z)$ as a programmable constructor (or automaton) $\vec{\Gamma}$ whose legitimate inputs are: the specification of the *game attribute* $z$, the (deterministic) *rules of the game* (with the game observable $\boldsymbol{X}$) and the subsidiary theory. Its program must also be such that it satisfy the following *axioms*:

A1. *Ordering*. Given $z$ and $z'$, the automaton orders any two games $G_X(z)$ and $G_X(z')$ - the ordering is transitive and total.

In this constructor-theoretic version of the decision-theory argument, this is the *only classical decision-theoretic axiom* required of the automaton. It corresponds to the *transitivity of preferences* in (Deutsch, 1999; Wallace, 2012). Its effect is to require $\vec{\Gamma}$ to be a constructor for the task of providing in output a *real number* $V\{G_X(z)\}\in X$ – the *value of the game.* Specifically, I define the *value* of the game $G_X(z)$, $V\{G_X(z)\}$, as the *unique* $v_z\in X$ with the property that $\vec{\Gamma}$ is indifferent between playing the game $G_X(z)$ and the game $G_X(v_z)$. As the reader may guess, the key will turn out to be that attributes with the same partition of unity have the same value.

A2. *Game of chance*. The only observables allowed to *condition*[24] the automaton's output are: (i) the observable for *whether the rules of the game are followed*; (ii) *the observable* $\boldsymbol{D}_{X_p}$, defined as the difference between the observable $\boldsymbol{X}_p$ of the first reward substrate before the game, and the observable $\boldsymbol{X}_p$ of the second reward substrate after the game, as predicted by the automaton (given the specification of the input attribute and the subsidiary theory).[25]

Thus, whether other observables than the ones of axiom 2 may be sharp is irrelevant to the automaton's output. Otherwise, those observables would have to be mentioned in the program to condition the automaton's output (thus specifying a player for a different game). This fact shall be repeatedly used in section 7.2, to show properties (P1-P6).

In classical decision theory any monotonic rescaling of the utility function causes no change in choices; so, without losing generality, I shall assume that whenever the observable $\boldsymbol{D}_{X_p}$ is predicted by the automaton to be *sharp on the payoff substrate with value x*, the automaton outputs a substrate holding a sharp $v_x = x$.

#### *7.2 Properties of the value function*

As I promised, axiom 1 and 2 imply *crucial properties*, which were construed as independent decision-theoretic axioms in earlier treatments, but here they follow from the other axioms and the principles of constructor theory, under decision-supporting superinformation theories. Namely:

---

[24] An observable $\boldsymbol{O}$ *conditions* the output of the automaton if its program, given the specification of the input attribute and the subsidiary theory, effectively contains an instruction: "if $\boldsymbol{O}$ is sharp with some value *o*, then …"

[25] Note that under constructor theory, as under quantum theory, this must be an observable because of the principle of consistency of measurement (section 3). (In quantum theory, those two observables commute even though they are at different times.)

P1. $\vec{\Gamma}$'s preferences must be *constant in time.* Otherwise the observable 'elapsed time' would have to condition the program, violating axiom A2. [26]

P2. *Substitutability of games* (Luce & Raiffa, 1957; Deutsch, 1999). The value of a game $G_X(\boldsymbol{z})$ in isolation must be the same as when composed with another game $G_X(\boldsymbol{z}')$. Otherwise, again, the automaton's program would have to be conditioned on the observable "what games $G_X(\boldsymbol{z})$ is composed with", violating axiom A2.

P3. *Additivity of composition.* Setting $V\{G_X(\boldsymbol{z})\}=v_z$ and $V\{G_X(\boldsymbol{z}')\}=v_{z'}$, P2 implies:

$$V\{G_X(\boldsymbol{z})\odot G_X(\boldsymbol{z}')\} \underset{\substack{\text{by substitutability, definition of value}}}{=} V\{G_X(\boldsymbol{v}_z)\odot G_X(\boldsymbol{v}_{z'})\} \underset{\substack{\text{by definition of } \odot,\\ \text{definition of value}}}{=} v_z+v_{z'} \tag{8}$$

P4. *Measurement neutrality* (Wallace, 2007). Games where the $X$-adder (under the given labellings) has *physically different implementations* must have the *same value* – otherwise the observable 'which physical implementation' should have to be included in the automaton's program, violating axiom A2. Measurement neutrality, in turn, allows one to show the following additional properties:

a. The game $G_X(\Pi(\boldsymbol{z}))$ – where the computation $\Pi$ (for any permutation $\Pi$ *over* $\boldsymbol{X}$) acts on **S** immediately before the adder (figure 6) – is a $\Pi(\boldsymbol{X})$-adder, where $\Pi(\boldsymbol{X})$ denotes the relabelling of $\boldsymbol{X}$ given by $\Pi$. Hence, it can be regarded as a particular physical implementation of the game $G_{\Pi(X)}(\boldsymbol{z})$. Therefore, measurement neutrality implies:

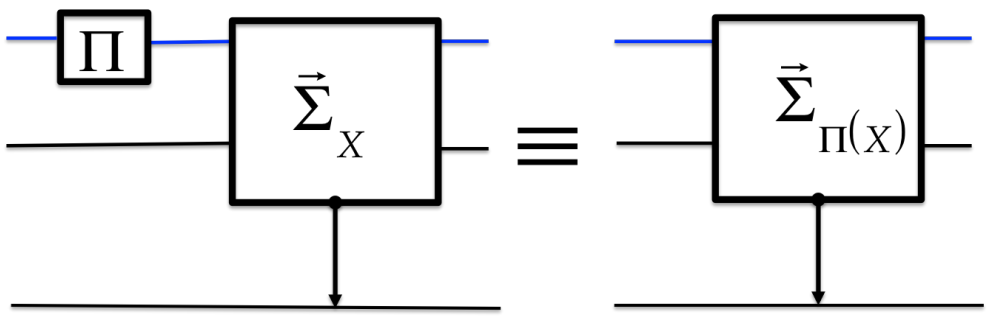

Figure 6

$$V\{G_{\Pi(X)}(\boldsymbol{z})\}=V\{G_X(\Pi(\boldsymbol{z}))\}\ . \tag{9}$$

[26] This condition is a cognate of Wallace's *diachronic consistency* (Wallace, 2012), although the latter is expressed via composition of games based on relative states, which here can be dispensed with.

b. Since $G_X(z_a)\odot G_X(z_b)$ is a particular physical implementation of $G_{X_a+X_b}((z_a,z_b))$ (figure 5), by measurement neutrality *the values of games on composite substrates must equal that of composite games*:

$$V\left\{G_{X_a+X_b}(z_a,z_b)\right\}=V\left\{G_X(z_a)\odot G_X(z_b)\right\}=v_{z_a}+v_{z_b} \tag{10}$$

where the last step follows from (8).

c. Let the attributes $[\boldsymbol{a}_y]_X$ and $[\boldsymbol{b}_y]_X$ be the $X$-intrinsic parts $[\boldsymbol{a}_y]_X\supseteq\boldsymbol{a}_y$ and $[\boldsymbol{b}_y]_X\supseteq\boldsymbol{b}_y$ of the attributes $\boldsymbol{a}_y$ and $\boldsymbol{b}_y$ (section 4), where $(\boldsymbol{a}_y,\boldsymbol{b}_y)$ is the attribute of $\mathbf{S}_a\oplus\mathbf{S}_b$ prepared by measuring $\boldsymbol{X}$ on $\mathbf{S}_a$ in the attribute $\boldsymbol{y}$, with $\mathbf{S}_b$ as a target. By prepending a measurer of $\boldsymbol{X}$ to the $X$-adder in the game $G_X(\boldsymbol{y})$, so that the measurer's source is the input of the $X$-adder (figure 7 bottom left), one still has the same $X$-adder. Thus, the $X$-game $G_X([\boldsymbol{a}_y]_X)$ (figure 7, top) has the same physical implementation as the $X$-game $G_X(\boldsymbol{y})$ (figure 7, bottom right). Similarly, by concatenating a measurer of $\boldsymbol{X}$ to an '$X$'-adder so that the former's target is the game substrate of the latter, one obtains an $X$-adder overall, whereby the '$X$'-game $G_{'X'}([\boldsymbol{b}_y]_X)$ has the same physical implementation as the $X$-game $G_X(\boldsymbol{y})$. The same applies if one considers $G_{\Pi(X)}(\boldsymbol{y})$, for any permutation $\Pi$ over $X$. Measurement neutrality implies:

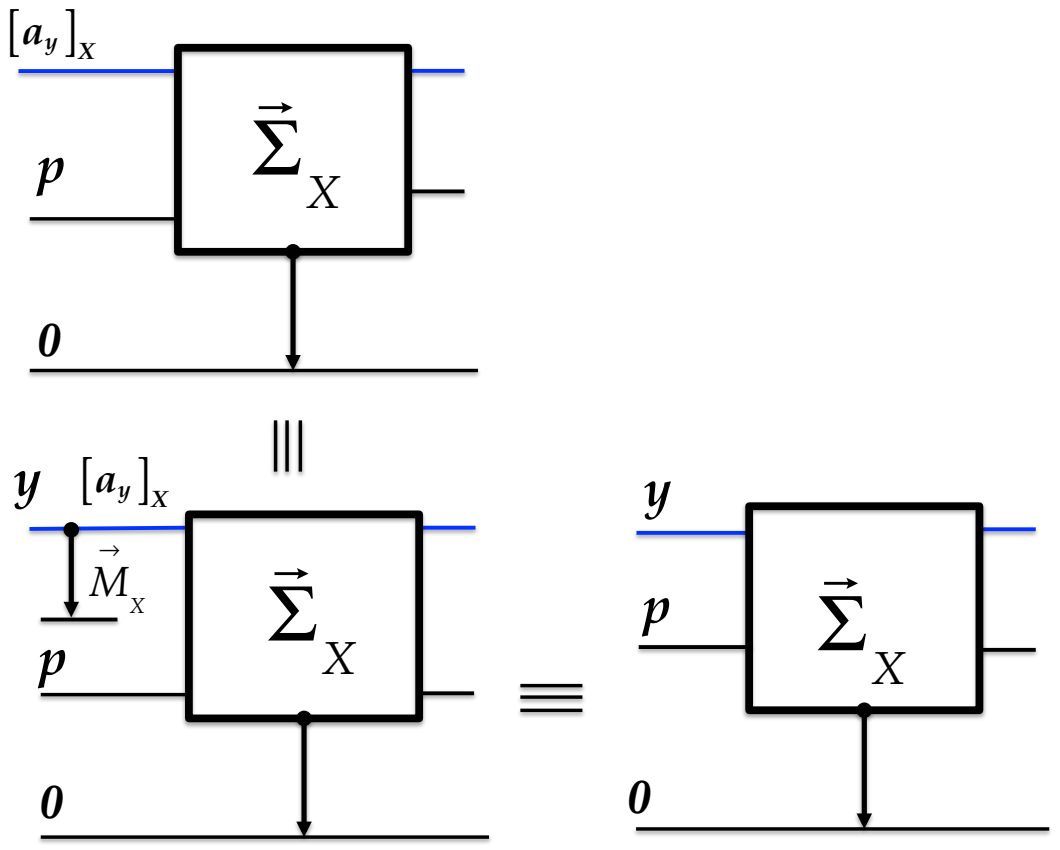

Figure 7

$$V\left\{G_{\Pi(X)}([\boldsymbol{a}_y]_X)\right\}=V\left\{G_{\Pi(X)}(\boldsymbol{y})\right\}=V\left\{G_{'\Pi(X)'}([\boldsymbol{b}_y]_X)\right\}. \tag{11}$$

d. *Shift rule (Deutsch 1999)*. Define the *uniform shift* as the permutation

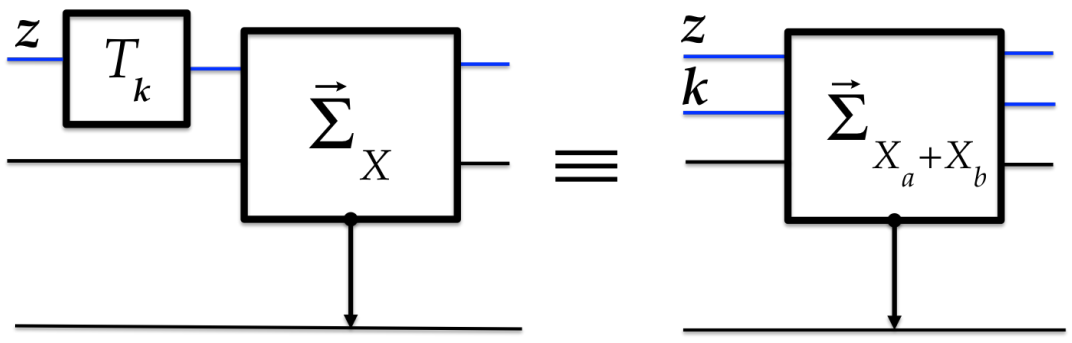

Figure 8

$T_k = \bigcup_{x \in X} \{x \to x + k\}$, for real *k*. The game $G_{X_a+X_b}(\boldsymbol{z},\boldsymbol{k})$ (figure 8, right) is a particular implementation of the game $G_{T_k(X)}(\boldsymbol{z})$ (figure 8, left), where the $T_k(X)$-adder is realised by measuring $\boldsymbol{X}_a + \boldsymbol{X}_b$ on $\mathbf{S}_a \oplus \mathbf{S}_b$ in the attribute $(\boldsymbol{z},\boldsymbol{k})$, for a fixed $\boldsymbol{k} \in \boldsymbol{X}$. By measurement neutrality and equation (10):

$$V\left\{G_{T_k(X)}(\boldsymbol{z})\right\} = V\left\{G_{X_a+X_b}(\boldsymbol{z},\boldsymbol{k})\right\} = v_z + k \ . \quad (12)$$

P5. *The "equal value" property*. In quantum theory, a superposition or mixture of two orthogonal quantum states such that *X*-games played with them have the same value *v*, has value *v* (Deutsch, 1999). A generalisation of this property holds under decision-supporting superinformation theories. A crucial difference from the argument in *(*Deutsch, 1999*)* is that here there is no need explicitly to resort to observed outcomes or relative states to prove this property. Let $\boldsymbol{H} = \{\boldsymbol{h}_1, \boldsymbol{h}_2\}$ be an information variable of a game substrate $\mathbf{S}_a$, such that *X*-games played with $\boldsymbol{h}_1$ or $\boldsymbol{h}_2$ *have the same value v*: $V\{G_X(\boldsymbol{h}_1)\} = v = V\{G_X(\boldsymbol{h}_2)\}$. Consider the information attribute $\boldsymbol{q}$, having an *X*-partition of unity, with the property that $\boldsymbol{q}$ is a generalised mixture of $\boldsymbol{H}$ – whereby $\boldsymbol{q} \subseteq \bar{\bar{\boldsymbol{u}}}_H$. I shall now prove that the *X*-game with game attribute $\boldsymbol{q}$ *has also value v*: $V\{G_X(\boldsymbol{q})\} = v$. In the trivial case that $\boldsymbol{q} = \boldsymbol{h}_1$ or $\boldsymbol{q} = \boldsymbol{h}_2$, $V\{G_X(\boldsymbol{q})\} = v$. Consider the case where, instead, $\boldsymbol{q}$ is a non-trivial generalised mixture of $\boldsymbol{H} = \{\boldsymbol{h}_1, \boldsymbol{h}_2\}$: $\boldsymbol{q} \subseteq \bar{\bar{\boldsymbol{u}}}_H$, $\boldsymbol{q} \cap \boldsymbol{h}_i = \emptyset$, $\boldsymbol{q} \not\perp \boldsymbol{h}_i, \forall i = 1,2$. In quantum theory, $\boldsymbol{q}$ corresponds to the attribute of being in a superposition or mixture of distinguishable quantum states $|h_1\rangle, |h_1\rangle$ : $\langle h_1 | h_2 \rangle = 0$ – e.g., the quantum state $|q\rangle = (\alpha|h_1\rangle + \beta|h_2\rangle)$ for complex $\alpha, \beta$. I shall show that $G_X(\boldsymbol{q})$ is equivalent to another game with value *v*, obtained by applying the following procedure on the game substrate $\mathbf{S}_a$.

First, measure $\boldsymbol{H}$ on $\mathbf{S}_a$ holding the attribute $\boldsymbol{q}$, with an ancillary system $\mathbf{S}_b$ as a target of the $\boldsymbol{H}$-measurer. This delivers the substrate $\mathbf{S}_a \oplus \mathbf{S}_b$ in the attribute $(\boldsymbol{a}_q, \boldsymbol{b}_q)$.

By the properties of the intrinsic part (section 3), $[\boldsymbol{a}_q]_X$ still is a generalized mixture of the attributes in ***H:*** $\boldsymbol{a}_q \subseteq [\boldsymbol{a}_q]_X \subseteq \bar{\bar{\boldsymbol{u}}}_H$, $[\boldsymbol{a}_q]_X \cap \boldsymbol{h}_i = \emptyset$, $[\boldsymbol{a}_q]_X \not\angle \boldsymbol{h}_i$, $(\forall i = 1,2)$. Thus, $V\{G_X(\boldsymbol{q})\} = V\{G_X([\boldsymbol{a}_q]_X)\}$, because both $\boldsymbol{q}$ and $[\boldsymbol{a}_q]_X$ have the same specification *qua* $X$-game attributes. Next, apply a constructor $\vec{C}_H$ for the following task (a control operation in quantum theory): if ***'H'*** is sharp with value $'\boldsymbol{h}_1'$ on $\mathbf{S}_b$, nothing happens on $\mathbf{S}_a$; whereas if ***'H'*** is sharp with value $'\boldsymbol{h}_2'$ on $\mathbf{S}_b$, then $\boldsymbol{h}_2$ on $\mathbf{S}_a$ is replaced by the attribute $\boldsymbol{h}_1$. Applying $\vec{C}_H$ does not change the value of the game on $\mathbf{S}_a$, for the attributes on the game substrate $\mathbf{S}_a$ before (i.e., $[\boldsymbol{a}_q]_X$) and after (i.e., $\boldsymbol{h}_1$) the action of $\vec{C}_H$ have the same specification: they are both generalised mixtures of attributes in ***H***. Thus, $V\{G_X(\boldsymbol{q})\} = V\{G_X([\boldsymbol{a}_q]_X)\} = V\{G_X(\boldsymbol{h}_1)\} = v$, as promised.

P6.*The reflection rule.*

Under decision-supporting superinformation theories, including quantum theory, the property that I shall call the *reflection rule* follows. It is the superinformation version of Deutsch's (1999) 'zero-sum rule', but, again, the present derivation does not depend on relative states or observed outcomes in universes. Consider the attributes $\boldsymbol{y}_+$ and $\boldsymbol{y}_-$, satisfying the condition R1-R4 (section 6). By (11) and R3 it follows that:

$$V\left\{G_{\Pi(X)}(\boldsymbol{y}_+)\right\} = V\left\{G_{\Pi(X)}([\boldsymbol{a}_{y_\pm}]_X)\right\} = V\left\{G_{\Pi(X)}(\boldsymbol{y}_-)\right\}$$

for any $\Pi$ over $X$. Hence, by the additivity of composition (property P3), it also follows that:

$$V\left\{G_{\Pi(X)_a+\Pi'(X)_b}(\boldsymbol{y}_+,\boldsymbol{y}_+)\right\} = V\left\{G_{\Pi(X)_a}(\boldsymbol{y}_+)\right\} + V\left\{G_{\Pi'(X)_b}(\boldsymbol{y}_+)\right\} = V\left\{G_{\Pi(X)_a+\Pi'(X)_b}(\boldsymbol{y}_-,\boldsymbol{y}_-)\right\} \quad (13)$$

for any permutations $\Pi$, $\Pi'$ over $X$. Consider the game $G_{R(X)_a+X_b}(\boldsymbol{q})$ where $\boldsymbol{q}$ satisfies conditions R4 (section 6) and the *reflection* $R$ is the permutation over $X$ defined by $R = \bigcup_{x \in X}\{x \to -x\}$. The adder in the game is a measurer of $\boldsymbol{R(X)}_a + \boldsymbol{X}_b$, which, in turn, is an $X$-comparer (as defined in equation (4)) on the substrates $\mathbf{S}_a$, $\mathbf{S}_b$: it measures the observable «whether the two substrates hold the same value $x$», where the output $\boldsymbol{0}$ corresponds to «***yes***». By the *consistency of successive measurements of* $\boldsymbol{X}$ on the *same* substrate with attribute $\boldsymbol{y}$, (section 3), the output variable of that measurer is sharp with value *0* when presented with an attribute which, like $\boldsymbol{q}$, has the property that $\boldsymbol{q} \subseteq \bar{\bar{u}}_{S_x}$ – where $S_x \doteq \{(\boldsymbol{x_1},\boldsymbol{x_1}),(\boldsymbol{x_2},\boldsymbol{x_2})\}$. In quantum theory, as I said, the attribute $\boldsymbol{q}$ corresponds to the attribute of being in the quantum state $\frac{1}{\sqrt{2}}\left[|x_1\rangle|x_1\rangle+|x_2\rangle|x_2\rangle\right]$, which is a *0*-eigenstate of the observable $-\hat{X}_a+\hat{X}_b$. Thus, by definition of value, the value of $G_{R(X)_a+X_b}(\boldsymbol{q})$ must be 0.

On the other hand, since $\boldsymbol{q} \subseteq \bar{\bar{u}}_{S_y}$ where $S_y \doteq \{(y_+,y_+),(y_-,y_-)\}$ and, by (13), both $G_{R(X)_a+X_b}((\boldsymbol{y}_-,\boldsymbol{y}_-))$ and $G_{R(X)_a+X_b}((\boldsymbol{y}_+,\boldsymbol{y}_+))$ have the *same value*, by the 'equal-value' property (e), one has: $V\left\{G_{R(X)_a+X_b}(\boldsymbol{q})\right\}=V\left\{G_{R(X)_a+X_b}((\boldsymbol{y}_+,\boldsymbol{y}_+))\right\}$

Hence:

$$0=V\left\{G_{R(X)_a+X_b}(\boldsymbol{q})\right\}=V\left\{G_{R(X)_a+X_b}((\boldsymbol{y}_+,\boldsymbol{y}_+))\right\}=V\left\{G_{R(X)}(\boldsymbol{y}_+)\odot G_X(\boldsymbol{y}_+)\right\}$$

where the last step follows by (10). Finally, one obtains, as promised:

$$V\left\{G_{R(X)}(\boldsymbol{y}_+)\right\}=-V\left\{G_X(\boldsymbol{y}_+)\right\} \quad (14)$$

It follows that if some attribute $\tilde{\boldsymbol{y}}$ is invariant under reflection ($R(\tilde{\boldsymbol{y}})\subseteq\tilde{\boldsymbol{y}}$) then $V\left\{G_X(\tilde{\boldsymbol{y}})\right\}=0$, because:

$$V\left\{G_X(\tilde{\boldsymbol{y}})\right\}=V\left\{G_X(R(\tilde{\boldsymbol{y}}))\right\} \overset{\text{Measurement Indifference}}{\overset{\downarrow}{=}} V\left\{G_{R(X)}(\tilde{\boldsymbol{y}})\right\} \overset{\text{Reflection Rule}}{\overset{\downarrow}{=}} -V\left\{G_X(\tilde{\boldsymbol{y}})\right\} \quad (15).$$

***7.3 The decision-theoretic argument under decision-supporting superinformation theories***

I shall now show that if an attribute $\boldsymbol{y}$ has $X$-partition of unity $\left[f(y)_{x_1},\ldots f(y)_{x_d}\right]$ then

$$V\left\{G_X(\boldsymbol{y})\right\}=\sum_{x\in X} f(y)_x x \quad (16)$$

thereby showing that a program satisfying the axioms in 7.1 is possible, under decision-supporting superinformation theories. The automaton with *that program* must value the game using the $f(y)_x$ *as if* they were the probabilities of outcomes, without assuming (or concluding!) that they are. Thus it places bets *in the same way as* a corresponding automaton would, if programmed with the same axioms of classical decision theory *plus additional, ad hoc probabilistic axioms* connecting probabilities with decisions (e.g., that the 'prudentially best option' maximises the expected value of the gain (Papineau, 2006)) *and* some stochastic theory with the $f(y)_x$ as probabilities of outcomes *x* (e.g. quantum theory with the Born Rule).

My argument broadly follows the logic of Deutsch's or Wallace's formulation; however, individual steps rest on different constructor-theoretic conditions and do not use concepts specific to quantum theory or the Everett interpretation – e.g., subspaces, observed outcomes, relative states, universes, and instances of the player in 'universes' or 'branches of the multiverse'.

Consider the attribute $\boldsymbol{y}_+$, satisfying conditions (R1-R4). Recall that it is symmetric under swap: $S_{x_1,x_2}(\boldsymbol{y}_+) \subseteq \boldsymbol{y}_+$, which implies $f(y_\pm)_{x_1} = f(y_\pm)_{x_2} = 1/2$. This special case of 'equal weights' provides the pivotal step in the decision-theory argument, just as in Deutsch or Wallace.

The shift rule (12) with $k = -(x_1 + x_2)$ gives:

$$V\{G_X(T_{-(x_1+x_2)}(\boldsymbol{y}_+))\} = V\{G_X(\boldsymbol{y}_+)\} - (x_1 + x_2) \qquad (17)$$

Now, $T_{-(x_1+x_2)}$ can be implemented by composing the swap $S_{x_1,x_2}$ and the reflection *R*. By measurement neutrality (P4), the reflection rule (14) and the symmetry $S_{x_1,x_2}(\boldsymbol{y}_+) \subseteq \boldsymbol{y}_+$:

$$V\left\{G_X(T_{-(x_1+x_2)}(\boldsymbol{y}_+))\right\} = V\left\{G_X(R(S_{x_1,x_2}(\boldsymbol{y}_+)))\right\} = -V\{G_X(\boldsymbol{y}_+)\}$$

By comparison with (17):

$$V\{G_X(\boldsymbol{y}_+)\} = \frac{(x_1 + x_2)}{2} \qquad (18)$$

which is a special case of (16) when the game attribute $\boldsymbol{y}_+$ has an $X$-partition of unity whose non-null elements are $f(\boldsymbol{y}_+)_{x_1} = f(\boldsymbol{y}_+)_{x_2} = 1/2$. This is the central result proved by Deutsch and Wallace, but now proved from the physical, constructor-theoretic properties of measurements. The general, 'unequal weights', case follows by an argument analogous to (Deutsch, 1999), with a few conceptual differences outlined in the appendix.

What elements of reality the elements of a partition of unity in (16) represent depends on the subsidiary theory in question: it is not up to constructor theory (nor to the decision-theory argument) to explain that. The decision-theoretic argument only shows that decisions can be made, when informed by a decision-supporting superinformation theory, under a *non-probabilistic* subset of the classical axioms of rationality, as if it were a stochastic theory with probabilities given by the elements in the partition of unity. In particular, in unitary quantum theory that argument is *not* (as it is often described) a 'derivation of the Born Rule', but an explanation of why, without assuming the Born Rule, in the situations where it would apply, one must use the moduli squared of the amplitudes (and only those) to inform decisions *as if* the Born Rule held. Here I have shown that the equivalent is true for any constructor-theory-compliant subsidiary theory satisfying the stated criteria. Thus, such theories can, just as unitary quantum theory, account for the appearance of stochastic behaviour without appealing to any stochastic law.

## 8. Conclusions

I have reformulated the problem of reconciling *unitary quantum theory*, *unpredictability* and the *appearance of stochasticity* in quantum systems, within the newly proposed *constructor theory of information,* where unitary quantum theory is a particular *superinformation theory* – a *non-probabilistic* theory obeying constructor-theoretic principles. I have provided an *exact criterion for unpredictability*; then I have shown that superinformation theories (and thus unitary quantum theory) satisfy that criterion, showing that unpredictability follows from the impossibility of cloning certain sets of states, and that it is compatible with deterministic laws. This distinguishes it from randomness.

Then, I have exhibited conditions under which superinformation theories can inform decisions in games of chance, *as if* they were stochastic theories – by giving conditions for *decision-supporting superinformation theories*. To this end, I have generalised to constructor theory the Deutsch-Wallace quantum decision-theoretic approach, which shows how unitary quantum theory (with no Born Rule) can inform decisions in those games *as if* the Born Rule were assumed.

My constructor-theoretic approach improves upon that approach in two ways: 1) Its axioms, formulated in constructor-information-theoretic terms *only*, make no use of concepts specific to Everettian quantum theory, thus broadening the domain of applicability of the approach to decision-supporting superinformation theories; 2) It shows that some of the assumptions that were previously considered as purely decision-theoretic, and thus criticised for being 'subjective' (viz. measurement neutrality, diachronic consistency, the zero-sum rule), follow from the *physical properties* of *superinformation media*, *measurers* and *adders,* as required by the newly proposed laws of constructor theory.

So the axioms of the decision-theory approach turn out *not* to be particular, ad-hoc axioms necessary for quantum theory only, as it was previously thought; instead, they are either physical, information-theoretic requirements (such as the principle of consistency of measurement, interoperability, and the conditions for decision-supporting superinformation theories) *or* general methodological rules of scientific methodology (such as *transitivity of preferences*) required by *general theory testing*. On this ground, Deutsch (2015) has recently shown how not only quantum theory, but all decision-supporting superinformation theories (as defined in this paper) are testable in regard to their statements about repeated unpredictable measurements.

The present paper and Deutsch's one, taken together, imply that it is possible to regard the set of *decision-supporting superinformation theories* as a set of theoretical possibilities for a local, non-probabilistic generalisation of quantum theory, thus providing a new framework where the successor of quantum theory may be sought.

## Acknowledgements

I thank Simon Benjamin for suggesting useful improvements; Alan Forrester for helpful comments; Vlatko Vedral and Andrew Garner for discussions about the Born Rule; David Wallace for discussions about Everettian quantum theory; and David Deutsch especially for sharp criticism and insightful comments on earlier versions of this paper. I also thank two anonymous referees for their helpful suggestions, and for pointing out several ways in which this manuscript could be improved. This publication was made possible through the support of a grant from the Templeton World Charity Foundation. The opinions expressed in this publication are those of the author and do not necessarily reflect the views of Templeton World Charity Foundation.

**Appendix**

To prove the general case one follows the same logic as in (Deutsch, 1999). There are a number of conceptual differences, which I shall outline here via an example from quantum theory.

One first addresses the *generalised symmetric case*, proving that

$$V\{G_X(\boldsymbol{y})\} = \frac{1}{n}\sum_{k=1}^{n} x_k \tag{19}$$

whenever $\boldsymbol{y} \subseteq \bar{\bar{\boldsymbol{u}}}_{X_y}$, $\boldsymbol{X}_y = \{\boldsymbol{x}_1, \boldsymbol{x}_2, ..., \boldsymbol{x}_n\}$ and $S_{x_i,x_j}(\boldsymbol{y}) \subseteq \boldsymbol{y}$ for all *i,j* - i.e., by the arguments in section 6.2, $\boldsymbol{y}$ has a partition of unity with elements $f(y)_{x_i} = 1/n, \ \forall \ x_i$. This corresponds to equally weighted superpositions or mixtures, e.g. $|y\rangle = \sqrt{\frac{1}{n}}\sum_{l=1}^{n}|x_l\rangle$. The proof follows directly from the equal-value property, P5, via the same steps as in (Deutsch, 1999), so it will not be reproduced here.

In the general *non-symmetric case*, one shows that for any integers $m \geq 0, \ n > m$:

$$V\{G_X(\boldsymbol{y})\} = \frac{mx_1 + (n-m)x_2}{n} \tag{20}$$

where $\boldsymbol{y}$ is a generalised mixture of $\boldsymbol{X}_y = \{\boldsymbol{x}_1, \boldsymbol{x}_2\}$, such that $f(y)_{x_1} = \frac{m}{n}$, so that (by the properties of the *X*-partition of unity) $f(y)_{x_2} = \frac{n-m}{n}$. In quantum theory, $\boldsymbol{y}$ corresponds to the attribute of being in some superposition or mixture with that partition of unity, e.g. $|y\rangle = \sqrt{\frac{m}{n}}|x_1\rangle + \sqrt{\frac{(n-m)}{n}}|x_2\rangle$. (The continuum case follows as in (Deutsch, 1999)).

From the additivity of composition (P3) and the definition of value, it follows that for any game attribute $\boldsymbol{y}$, $V\{G_{X_a+X_b}((\boldsymbol{y},\tilde{\boldsymbol{y}}))\} = V\{G_X(\boldsymbol{y})\}$, whenever $V\{G_X(\tilde{\boldsymbol{y}})\} = 0$.

One way of preparing the composite substrate $\mathbf{S}_a \oplus \mathbf{S}_b$ so that $\mathbf{S}_a$ has an attribute with the same value as $\boldsymbol{y}$ and $\mathbf{S}_b$ holds an attribute with value 0 is to prepare the quantum state $|s\rangle = \left(\sqrt{m/n}\,|x_1\rangle|o_1\rangle + \sqrt{(n-m)/n}\,|x_2\rangle|o_2\rangle\right)$ on $\mathbf{S}_a \oplus \mathbf{S}_b$, by applying an *X*-measurer to $\mathbf{S}_a$ in the state $|y\rangle$, such that the measurer's output attribute $\boldsymbol{o}_1$ is that of being in the state $|o_1\rangle \doteq \frac{1}{\sqrt{m}}\sum_{n=1}|x_n\rangle$, and the output attribute $\boldsymbol{o}_2$ is that of being in the state $|o_2\rangle \doteq \frac{1}{\sqrt{(n-m)}} \sum_{n=m+1} |x_n\rangle$, *where both attributes are chosen to be invariant under reflection,* $R|o_i\rangle = |o_i\rangle$, and distinguishable from one another, $\langle o_1|o_2\rangle = 0$.[27] By equation (15) it follows that $V\{G_X(\boldsymbol{o}_i)\} = 0$. When $\mathbf{S}_a \oplus \mathbf{S}_b$ has the attribute $\boldsymbol{s} = (\boldsymbol{a}_y, \boldsymbol{b}_y)$ of being in the quantum state $|s\rangle$, the substrate $\mathbf{S}_a$ holds the game attribute $\left[\boldsymbol{a}_y\right]_X$ – the $X$-intrinsic part of $\boldsymbol{a}_y$. The substrate $\mathbf{S}_b$ has the game attribute $\left[\boldsymbol{b}_y\right]_{'X'}$ corresponding to the '$X$'-intrinsic part of $\boldsymbol{b}_y$, which, by the properties of the intrinsic part, is a generalised mixture of $\boldsymbol{o}_1$ and $\boldsymbol{o}_2$, each of which generates an $X$-game with value 0. Because of the equal-value property, the game played with $\left[\boldsymbol{b}_y\right]_{'X'}$ must have the same value 0. Therefore

$$V\left\{G_{X_a+X_b}(\boldsymbol{s})\right\} = V\left\{G_X(\left[\boldsymbol{a}_y\right]_X)\right\} + V\left\{G_X(\left[\boldsymbol{b}_y\right]_{'X'})\right\} = V\left\{G_X(\left[\boldsymbol{a}_y\right]_X)\right\} = V\left\{G_X(\boldsymbol{y})\right\} \tag{21}$$

where the last step follows from the property c.

On the other hand, since $|s\rangle = \frac{1}{\sqrt{n}}\left(|x_1\rangle \sum_{n=1}^{m} |x_n\rangle + |x_2\rangle \sum_{n=m+1}^{n} |x_n\rangle\right)$, the symmetric-case result (18) applies, giving $V\left\{G_{X_a+X_b}(\boldsymbol{s})\right\} = \frac{mx_1 + (n-m)x_2}{n}$, as $X_a + X_b$ is sharp in the attribute of being in the state $\sum_{k=1}^{m} |x_1\rangle |x_k\rangle$ with value $mx_2$ and in that of being in $\sum_{k=m+1}^{n} |x_1\rangle |x_k\rangle$ with value $(n-m)x_2$. Thus equation (21) proves the result (20):

$$V\left\{G_X(\boldsymbol{y})\right\} = \frac{mx_1 + (n-m)x_2}{n},$$

as promised.

[27] That this preparation is possible is guaranteed by the properties of quantum theory, but it is not clear whether that is true in general superinformation theories; I conjecture that it is, but if not, it would an additional *physical* condition on decision-supporting information theories, to be included in table 1.